\newif\ifams
\newif\ifmscs
\amstrue
\ifams
\documentclass[reqno,11pt]{amsart}
\usepackage[foot]{amsaddr}
\fi
\ifmscs
\documentclass[11pt]{mscs}
\fi
\usepackage{tikz}
\usepackage{tikz-cd}
\usetikzlibrary{calc}
\usepackage{amsmath}
\usepackage{amssymb}
\usepackage{amstext}
\usepackage{amsfonts}
\usepackage{xspace}
\usepackage{url}
\usepackage{enumerate}

\input{Macro}
\long\def\ig#1{\relax}
\ig{Thanks to Roberto Minio for this def'n.  Compare the def'n of
\comment in AMSTeX.}
 
\newcount \coefa
\newcount \coefb
\newcount \coefc
\newcount\tempcounta
\newcount\tempcountb
\newcount\tempcountc
\newcount\tempcountd
\newcount\xext
\newcount\yext
\newcount\xoff
\newcount\yoff
\newcount\gap%
\newcount\arrowtypea
\newcount\arrowtypeb
\newcount\arrowtypec
\newcount\arrowtyped
\newcount\arrowtypee
\newcount\height
\newcount\width
\newcount\xpos
\newcount\ypos
\newcount\run
\newcount\rise
\newcount\arrowlength
\newcount\halflength
\newcount\arrowtype
\newdimen\tempdimen
\newdimen\xlen
\newdimen\ylen
\newsavebox{\tempboxa}%
\newsavebox{\tempboxb}%
\newsavebox{\tempboxc}%
 
\makeatletter
\def\settypes(#1,#2,#3){\arrowtypea#1 \arrowtypeb#2 \arrowtypec#3}
\def\settoheight#1#2{\setbox\@tempboxa\hbox{#2}#1\ht\@tempboxa\relax}%
\def\settodepth#1#2{\setbox\@tempboxa\hbox{#2}#1\dp\@tempboxa\relax}%
\def\settokens[#1`#2`#3`#4]{%
     \def\tokena{#1}\def\tokenb{#2}\def\tokenc{#3}\def\tokend{#4}}
\def\setsqparms[#1`#2`#3`#4;#5`#6]{%
\arrowtypea #1
\arrowtypeb #2
\arrowtypec #3
\arrowtyped #4
\width #5
\height #6
}
\def\setpos(#1,#2){\xpos=#1 \ypos#2}
 
\def\bfig{\begin{picture}(\xext,\yext)(\xoff,\yoff)}
\def\efig{\end{picture}}
 
\def\putbox(#1,#2)#3{\put(#1,#2){\makebox(0,0){$#3$}}}
 
\def\settriparms[#1`#2`#3;#4]{\settripairparms[#1`#2`#3`1`1;#4]}%
 
\def\settripairparms[#1`#2`#3`#4`#5;#6]{%
\arrowtypea #1
\arrowtypeb #2
\arrowtypec #3
\arrowtyped #4
\arrowtypee #5
\width #6
\height #6
}
 
\def\resetparms{\settripairparms[1`1`1`1`1;500]\width 500}
 
\resetparms
 
\def\mvector(#1,#2)#3{
\put(0,0){\vector(#1,#2){#3}}%
\put(0,0){\vector(#1,#2){30}}%
}
\def\evector(#1,#2)#3{{
\arrowlength #3
\put(0,0){\vector(#1,#2){\arrowlength}}%
\advance \arrowlength by-30
\put(0,0){\vector(#1,#2){\arrowlength}}%
}}
 
\def\horsize#1#2{%
\settowidth{\tempdimen}{$#2$}%
#1=\tempdimen
\divide #1 by\unitlength
}
 
\def\vertsize#1#2{%
\settoheight{\tempdimen}{$#2$}%
#1=\tempdimen
\settodepth{\tempdimen}{$#2$}%
\advance #1 by\tempdimen
\divide #1 by\unitlength
}
 
\def\vertadjust[#1`#2`#3]{%
\vertsize{\tempcounta}{#1}%
\vertsize{\tempcountb}{#2}%
\ifnum \tempcounta<\tempcountb \tempcounta=\tempcountb \fi
\divide\tempcounta by2
\vertsize{\tempcountb}{#3}%
\ifnum \tempcountb>0 \advance \tempcountb by20 \fi
\ifnum \tempcounta<\tempcountb \tempcounta=\tempcountb \fi
}
 
\def\horadjust[#1`#2`#3]{%
\horsize{\tempcounta}{#1}%
\horsize{\tempcountb}{#2}%
\ifnum \tempcounta<\tempcountb \tempcounta=\tempcountb \fi
\divide\tempcounta by20
\horsize{\tempcountb}{#3}%
\ifnum \tempcountb>0 \advance \tempcountb by60 \fi
\ifnum \tempcounta<\tempcountb \tempcounta=\tempcountb \fi
}
 
\ig{ In this procedure, #1 is the paramater that sticks out all the way,
#2 sticks out the least and #3 is a label sticking out half way.  #4 is
the amount of the offset.}
 
\def\sladjust[#1`#2`#3]#4{%
\tempcountc=#4
\horsize{\tempcounta}{#1}%
\divide \tempcounta by2
\horsize{\tempcountb}{#2}%
\divide \tempcountb by2
\advance \tempcountb by-\tempcountc
\ifnum \tempcounta<\tempcountb \tempcounta=\tempcountb\fi
\divide \tempcountc by2
\horsize{\tempcountb}{#3}%
\advance \tempcountb by-\tempcountc
\ifnum \tempcountb>0 \advance \tempcountb by80\fi
\ifnum \tempcounta<\tempcountb \tempcounta=\tempcountb\fi
\advance\tempcounta by20
}
 
\def\putvector(#1,#2)(#3,#4)#5#6{{%
\xpos=#1
\ypos=#2
\run=#3
\rise=#4
\arrowlength=#5
\arrowtype=#6
\ifnum \arrowtype<0
    \ifnum \run=0
        \advance \ypos by-\arrowlength
    \else
        \tempcounta \arrowlength
        \multiply \tempcounta by\rise
        \divide \tempcounta by\run
        \ifnum\run>0
            \advance \xpos by\arrowlength
            \advance \ypos by\tempcounta
        \else
            \advance \xpos by-\arrowlength
            \advance \ypos by-\tempcounta
        \fi
    \fi
    \multiply \arrowtype by-1
    \multiply \rise by-1
    \multiply \run by-1
\fi
\ifnum \arrowtype=1
    \put(\xpos,\ypos){\vector(\run,\rise){\arrowlength}}%
\else\ifnum \arrowtype=2
    \put(\xpos,\ypos){\mvector(\run,\rise)\arrowlength}%
\else\ifnum\arrowtype=3
    \put(\xpos,\ypos){\evector(\run,\rise){\arrowlength}}%
\fi\fi\fi
}}
 
\def\putsplitvector(#1,#2)#3#4{
\xpos #1
\ypos #2
\arrowtype #4
\halflength #3
\arrowlength #3
\gap 140
\advance \halflength by-\gap
\divide \halflength by2
\ifnum \arrowtype=1
    \put(\xpos,\ypos){\line(0,-1){\halflength}}%
    \advance\ypos by-\halflength
    \advance\ypos by-\gap
    \put(\xpos,\ypos){\vector(0,-1){\halflength}}%
\else\ifnum \arrowtype=2
    \put(\xpos,\ypos){\line(0,-1)\halflength}%
    \put(\xpos,\ypos){\vector(0,-1)3}%
    \advance\ypos by-\halflength
    \advance\ypos by-\gap
    \put(\xpos,\ypos){\vector(0,-1){\halflength}}%
\else\ifnum\arrowtype=3
    \put(\xpos,\ypos){\line(0,-1)\halflength}%
    \advance\ypos by-\halflength
    \advance\ypos by-\gap
    \put(\xpos,\ypos){\evector(0,-1){\halflength}}%
\else\ifnum \arrowtype=-1
    \advance \ypos by-\arrowlength
    \put(\xpos,\ypos){\line(0,1){\halflength}}%
    \advance\ypos by\halflength
    \advance\ypos by\gap
    \put(\xpos,\ypos){\vector(0,1){\halflength}}%
\else\ifnum \arrowtype=-2
    \advance \ypos by-\arrowlength
    \put(\xpos,\ypos){\line(0,1)\halflength}%
    \put(\xpos,\ypos){\vector(0,1)3}%
    \advance\ypos by\halflength
    \advance\ypos by\gap
    \put(\xpos,\ypos){\vector(0,1){\halflength}}%
\else\ifnum\arrowtype=-3
    \advance \ypos by-\arrowlength
    \put(\xpos,\ypos){\line(0,1)\halflength}%
    \advance\ypos by\halflength
    \advance\ypos by\gap
    \put(\xpos,\ypos){\evector(0,1){\halflength}}%
\fi\fi\fi\fi\fi\fi
}
 
\def\putmorphism(#1)(#2,#3)[#4`#5`#6]#7#8#9{{%
\run #2
\rise #3
\ifnum\rise=0
  \puthmorphism(#1)[#4`#5`#6]{#7}{#8}{#9}%
\else\ifnum\run=0
  \putvmorphism(#1)[#4`#5`#6]{#7}{#8}{#9}%
\else
\setpos(#1)%
\arrowlength #7
\arrowtype #8
\ifnum\run=0
\else\ifnum\rise=0
\else
\ifnum\run>0
    \coefa=1
\else
   \coefa=-1
\fi
\ifnum\arrowtype>0
   \coefb=0
   \coefc=-1
\else
   \coefb=\coefa
   \coefc=1
   \arrowtype=-\arrowtype
\fi
\width=2
\multiply \width by\run
\divide \width by\rise
\ifnum \width<0  \width=-\width\fi
\advance\width by60
\if l#9 \width=-\width\fi
\putbox(\xpos,\ypos){#4}
{\multiply \coefa by\arrowlength
\advance\xpos by\coefa
\multiply \coefa by\rise
\divide \coefa by\run
\advance \ypos by\coefa
\putbox(\xpos,\ypos){#5} }%
{\multiply \coefa by\arrowlength
\divide \coefa by2
\advance \xpos by\coefa
\advance \xpos by\width
\multiply \coefa by\rise
\divide \coefa by\run
\advance \ypos by\coefa
\if l#9%
   \put(\xpos,\ypos){\makebox(0,0)[r]{$#6$}}%
\else\if r#9%
   \put(\xpos,\ypos){\makebox(0,0)[l]{$#6$}}%
\fi\fi }%
{\multiply \rise by-\coefc
\multiply \run by-\coefc
\multiply \coefb by\arrowlength
\advance \xpos by\coefb
\multiply \coefb by\rise
\divide \coefb by\run
\advance \ypos by\coefb
\multiply \coefc by70
\advance \ypos by\coefc
\multiply \coefc by\run
\divide \coefc by\rise
\advance \xpos by\coefc
\multiply \coefa by140
\multiply \coefa by\run
\divide \coefa by\rise
\advance \arrowlength by\coefa
\ifnum \arrowtype=1
   \put(\xpos,\ypos){\vector(\run,\rise){\arrowlength}}%
\else\ifnum\arrowtype=2
   \put(\xpos,\ypos){\mvector(\run,\rise){\arrowlength}}%
\else\ifnum\arrowtype=3
   \put(\xpos,\ypos){\evector(\run,\rise){\arrowlength}}%
\fi\fi\fi}\fi\fi\fi\fi}}
 
\def\puthmorphism(#1,#2)[#3`#4`#5]#6#7#8{{%
\xpos #1
\ypos #2
\width #6
\arrowlength #6
\putbox(\xpos,\ypos){#3\vphantom{#4}}%
{\advance \xpos by\arrowlength
\putbox(\xpos,\ypos){\vphantom{#3}#4}}%
\horsize{\tempcounta}{#3}%
\horsize{\tempcountb}{#4}%
\divide \tempcounta by2
\divide \tempcountb by2
\advance \tempcounta by30
\advance \tempcountb by30
\advance \xpos by\tempcounta
\advance \arrowlength by-\tempcounta
\advance \arrowlength by-\tempcountb
\putvector(\xpos,\ypos)(1,0){\arrowlength}{#7}%
\divide \arrowlength by2
\advance \xpos by\arrowlength
\vertsize{\tempcounta}{#5}%
\divide\tempcounta by2
\advance \tempcounta by20
\if a#8 %
   \advance \ypos by\tempcounta
   \putbox(\xpos,\ypos){#5}%
\else
   \advance \ypos by-\tempcounta
   \putbox(\xpos,\ypos){#5}%
\fi}}
 
\def\putvmorphism(#1,#2)[#3`#4`#5]#6#7#8{{%
\xpos #1
\ypos #2
\arrowlength #6
\arrowtype #7
\settowidth{\xlen}{$#5$}%
\putbox(\xpos,\ypos){#3}%
{\advance \ypos by-\arrowlength
\putbox(\xpos,\ypos){#4}}%
{\advance\arrowlength by-140
\advance \ypos by-70
\ifdim\xlen>0pt
   \if m#8%
      \putsplitvector(\xpos,\ypos){\arrowlength}{\arrowtype}%
   \else
      \putvector(\xpos,\ypos)(0,-1){\arrowlength}{\arrowtype}%
   \fi
\else
   \putvector(\xpos,\ypos)(0,-1){\arrowlength}{\arrowtype}%
\fi}%
\ifdim\xlen>0pt
   \divide \arrowlength by2
   \advance\ypos by-\arrowlength
   \if l#8%
      \advance \xpos by-40
      \put(\xpos,\ypos){\makebox(0,0)[r]{$#5$}}%
   \else\if r#8%
      \advance \xpos by40
      \put(\xpos,\ypos){\makebox(0,0)[l]{$#5$}}%
   \else
      \putbox(\xpos,\ypos){#5}%
   \fi\fi
\fi
}}
 
\def\topadjust[#1`#2`#3]{%
\yoff=10
\vertadjust[#1`#2`{#3}]%
\advance \yext by\tempcounta
\advance \yext by 10
}
\def\botadjust[#1`#2`#3]{%
\vertadjust[#1`#2`{#3}]%
\advance \yext by\tempcounta
\advance \yoff by-\tempcounta
}
\def\leftadjust[#1`#2`#3]{%
\xoff=0
\horadjust[#1`#2`{#3}]%
\advance \xext by\tempcounta
\advance \xoff by-\tempcounta
}
\def\rightadjust[#1`#2`#3]{%
\horadjust[#1`#2`{#3}]%
\advance \xext by\tempcounta
}
\def\rightsladjust[#1`#2`#3]{%
\sladjust[#1`#2`{#3}]{\width}%
\advance \xext by\tempcounta
}
\def\leftsladjust[#1`#2`#3]{%
\xoff=0
\sladjust[#1`#2`{#3}]{\width}%
\advance \xext by\tempcounta
\advance \xoff by-\tempcounta
}
\def\adjust[#1`#2;#3`#4;#5`#6;#7`#8]{%
\topadjust[#1``{#2}]
\leftadjust[#3``{#4}]
\rightadjust[#5``{#6}]
\botadjust[#7``{#8}]}
 
\def\putsquarep<#1>(#2)[#3;#4`#5`#6`#7]{{%
\setsqparms[#1]%
\setpos(#2)%
\settokens[#3]%
\puthmorphism(\xpos,\ypos)[\tokenc`\tokend`{#7}]{\width}{\arrowtyped}b%
\advance\ypos by \height
\puthmorphism(\xpos,\ypos)[\tokena`\tokenb`{#4}]{\width}{\arrowtypea}a%
\putvmorphism(\xpos,\ypos)[``{#5}]{\height}{\arrowtypeb}l%
\advance\xpos by \width
\putvmorphism(\xpos,\ypos)[``{#6}]{\height}{\arrowtypec}r%
}}
 
\def\putsquare{\@ifnextchar <{\putsquarep}{\putsquarep%
   <\arrowtypea`\arrowtypeb`\arrowtypec`\arrowtyped;\width`\height>}}
\def\square{\@ifnextchar< {\squarep}{\squarep
   <\arrowtypea`\arrowtypeb`\arrowtypec`\arrowtyped;\width`\height>}}
\def\squarep<#1>[#2`#3`#4`#5;#6`#7`#8`#9]{{
\setsqparms[#1]
\xext=\width                                          
\yext=\height                                         
\topadjust[#2`#3`{#6}]
\botadjust[#4`#5`{#9}]
\leftadjust[#2`#4`{#7}]
\rightadjust[#3`#5`{#8}]
\begin{picture}(\xext,\yext)(\xoff,\yoff)
\putsquarep<\arrowtypea`\arrowtypeb`\arrowtypec`\arrowtyped;\width`\height>%
(0,0)[#2`#3`#4`#5;#6`#7`#8`{#9}]%
\end{picture}%
}}
 
\def\putptrianglep<#1>(#2,#3)[#4`#5`#6;#7`#8`#9]{{%
\settriparms[#1]%
\xpos=#2 \ypos=#3
\advance\ypos by \height
\puthmorphism(\xpos,\ypos)[#4`#5`{#7}]{\height}{\arrowtypea}a%
\putvmorphism(\xpos,\ypos)[`#6`{#8}]{\height}{\arrowtypeb}l%
\advance\xpos by\height
\putmorphism(\xpos,\ypos)(-1,-1)[``{#9}]{\height}{\arrowtypec}r%
}}
 
\def\putptriangle{\@ifnextchar <{\putptrianglep}{\putptrianglep
   <\arrowtypea`\arrowtypeb`\arrowtypec;\height>}}
\def\ptriangle{\@ifnextchar <{\ptrianglep}{\ptrianglep
   <\arrowtypea`\arrowtypeb`\arrowtypec;\height>}}
 
\def\ptrianglep<#1>[#2`#3`#4;#5`#6`#7]{{
\settriparms[#1]%
\width=\height                         
\xext=\width                           
\yext=\width                           
\topadjust[#2`#3`{#5}]
\botadjust[#3``]
\leftadjust[#2`#4`{#6}]
\rightsladjust[#3`#4`{#7}]
\begin{picture}(\xext,\yext)(\xoff,\yoff)
\putptrianglep<\arrowtypea`\arrowtypeb`\arrowtypec;\height>%
(0,0)[#2`#3`#4;#5`#6`{#7}]%
\end{picture}%
}}
 
\def\putqtrianglep<#1>(#2,#3)[#4`#5`#6;#7`#8`#9]{{%
\settriparms[#1]%
\xpos=#2 \ypos=#3
\advance\ypos by\height
\puthmorphism(\xpos,\ypos)[#4`#5`{#7}]{\height}{\arrowtypea}a%
\putmorphism(\xpos,\ypos)(1,-1)[``{#8}]{\height}{\arrowtypeb}l%
\advance\xpos by\height
\putvmorphism(\xpos,\ypos)[`#6`{#9}]{\height}{\arrowtypec}r%
}}
 
\def\putqtriangle{\@ifnextchar <{\putqtrianglep}{\putqtrianglep
   <\arrowtypea`\arrowtypeb`\arrowtypec;\height>}}
\def\qtriangle{\@ifnextchar <{\qtrianglep}{\qtrianglep
   <\arrowtypea`\arrowtypeb`\arrowtypec;\height>}}
 
\def\qtrianglep<#1>[#2`#3`#4;#5`#6`#7]{{
\settriparms[#1]
\width=\height                         
\xext=\width                           
\yext=\height                          
\topadjust[#2`#3`{#5}]
\botadjust[#4``]
\leftsladjust[#2`#4`{#6}]
\rightadjust[#3`#4`{#7}]
\begin{picture}(\xext,\yext)(\xoff,\yoff)
\putqtrianglep<\arrowtypea`\arrowtypeb`\arrowtypec;\height>%
(0,0)[#2`#3`#4;#5`#6`{#7}]%
\end{picture}%
}}
 
\def\putdtrianglep<#1>(#2,#3)[#4`#5`#6;#7`#8`#9]{{%
\settriparms[#1]%
\xpos=#2 \ypos=#3
\puthmorphism(\xpos,\ypos)[#5`#6`{#9}]{\height}{\arrowtypec}b%
\advance\xpos by \height \advance\ypos by\height
\putmorphism(\xpos,\ypos)(-1,-1)[``{#7}]{\height}{\arrowtypea}l%
\putvmorphism(\xpos,\ypos)[#4``{#8}]{\height}{\arrowtypeb}r%
}}
 
\def\putdtriangle{\@ifnextchar <{\putdtrianglep}{\putdtrianglep
   <\arrowtypea`\arrowtypeb`\arrowtypec;\height>}}
\def\dtriangle{\@ifnextchar <{\dtrianglep}{\dtrianglep
   <\arrowtypea`\arrowtypeb`\arrowtypec;\height>}}
 
\def\dtrianglep<#1>[#2`#3`#4;#5`#6`#7]{{
\settriparms[#1]
\width=\height                         
\xext=\width                           
\yext=\height                          
\topadjust[#2``]
\botadjust[#3`#4`{#7}]
\leftsladjust[#3`#2`{#5}]
\rightadjust[#2`#4`{#6}]
\begin{picture}(\xext,\yext)(\xoff,\yoff)
\putdtrianglep<\arrowtypea`\arrowtypeb`\arrowtypec;\height>%
(0,0)[#2`#3`#4;#5`#6`{#7}]%
\end{picture}%
}}
 
\def\putbtrianglep<#1>(#2,#3)[#4`#5`#6;#7`#8`#9]{{%
\settriparms[#1]%
\xpos=#2 \ypos=#3
\puthmorphism(\xpos,\ypos)[#5`#6`{#9}]{\height}{\arrowtypec}b%
\advance\ypos by\height
\putmorphism(\xpos,\ypos)(1,-1)[``{#8}]{\height}{\arrowtypeb}r%
\putvmorphism(\xpos,\ypos)[#4``{#7}]{\height}{\arrowtypea}l%
}}
 
\def\putbtriangle{\@ifnextchar <{\putbtrianglep}{\putbtrianglep
   <\arrowtypea`\arrowtypeb`\arrowtypec;\height>}}
\def\btriangle{\@ifnextchar <{\btrianglep}{\btrianglep
   <\arrowtypea`\arrowtypeb`\arrowtypec;\height>}}
 
\def\btrianglep<#1>[#2`#3`#4;#5`#6`#7]{{
\settriparms[#1]
\width=\height                         
\xext=\width                           
\yext=\height                          
\topadjust[#2``]
\botadjust[#3`#4`{#7}]
\leftadjust[#2`#3`{#5}]
\rightsladjust[#4`#2`{#6}]
\begin{picture}(\xext,\yext)(\xoff,\yoff)
\putbtrianglep<\arrowtypea`\arrowtypeb`\arrowtypec;\height>%
(0,0)[#2`#3`#4;#5`#6`{#7}]%
\end{picture}%
}}
 
\def\putAtrianglep<#1>(#2,#3)[#4`#5`#6;#7`#8`#9]{{%
\settriparms[#1]%
\xpos=#2 \ypos=#3
{\multiply \height by2
\puthmorphism(\xpos,\ypos)[#5`#6`{#9}]{\height}{\arrowtypec}b}%
\advance\xpos by\height \advance\ypos by\height
\putmorphism(\xpos,\ypos)(-1,-1)[#4``{#7}]{\height}{\arrowtypea}l%
\putmorphism(\xpos,\ypos)(1,-1)[``{#8}]{\height}{\arrowtypeb}r%
}}
 
\def\putAtriangle{\@ifnextchar <{\putAtrianglep}{\putAtrianglep
   <\arrowtypea`\arrowtypeb`\arrowtypec;\height>}}
\def\Atriangle{\@ifnextchar <{\Atrianglep}{\Atrianglep
   <\arrowtypea`\arrowtypeb`\arrowtypec;\height>}}
 
\def\Atrianglep<#1>[#2`#3`#4;#5`#6`#7]{{
\settriparms[#1]
\width=\height                         
\xext=\width                           
\yext=\height                          
\topadjust[#2``]
\botadjust[#3`#4`{#7}]
\multiply \xext by2 
\leftsladjust[#3`#2`{#5}]
\rightsladjust[#4`#2`{#6}]
\begin{picture}(\xext,\yext)(\xoff,\yoff)%
\putAtrianglep<\arrowtypea`\arrowtypeb`\arrowtypec;\height>%
(0,0)[#2`#3`#4;#5`#6`{#7}]%
\end{picture}%
}}
 
\def\putAtrianglepairp<#1>(#2)[#3;#4`#5`#6`#7`#8]{{
\settripairparms[#1]%
\setpos(#2)%
\settokens[#3]%
\puthmorphism(\xpos,\ypos)[\tokenb`\tokenc`{#7}]{\height}{\arrowtyped}b%
\advance\xpos by\height
\advance\ypos by\height
\putmorphism(\xpos,\ypos)(-1,-1)[\tokena``{#4}]{\height}{\arrowtypea}l%
\putvmorphism(\xpos,\ypos)[``{#5}]{\height}{\arrowtypeb}m%
\putmorphism(\xpos,\ypos)(1,-1)[``{#6}]{\height}{\arrowtypec}r%
}}
 
\def\putAtrianglepair{\@ifnextchar <{\putAtrianglepairp}{\putAtrianglepairp%
   <\arrowtypea`\arrowtypeb`\arrowtypec`\arrowtyped`\arrowtypee;\height>}}
\def\Atrianglepair{\@ifnextchar <{\Atrianglepairp}{\Atrianglepairp%
   <\arrowtypea`\arrowtypeb`\arrowtypec`\arrowtyped`\arrowtypee;\height>}}
 
\def\Atrianglepairp<#1>[#2;#3`#4`#5`#6`#7]{{%
\settripairparms[#1]%
\settokens[#2]%
\width=\height
\xext=\width
\yext=\height
\topadjust[\tokena``]%
\vertadjust[\tokenb`\tokenc`{#6}]
\tempcountd=\tempcounta                       
\vertadjust[\tokenc`\tokend`{#7}]
\ifnum\tempcounta<\tempcountd                 
\tempcounta=\tempcountd\fi                    
\advance \yext by\tempcounta                  
\advance \yoff by-\tempcounta                 %
\multiply \xext by2 
\leftsladjust[\tokenb`\tokena`{#3}]
\rightsladjust[\tokend`\tokena`{#5}]%
\begin{picture}(\xext,\yext)(\xoff,\yoff)%
\putAtrianglepairp
<\arrowtypea`\arrowtypeb`\arrowtypec`\arrowtyped`\arrowtypee;\height>%
(0,0)[#2;#3`#4`#5`#6`{#7}]%
\end{picture}%
}}
 
\def\putVtrianglep<#1>(#2,#3)[#4`#5`#6;#7`#8`#9]{{%
\settriparms[#1]%
\xpos=#2 \ypos=#3
\advance\ypos by\height
{\multiply\height by2
\puthmorphism(\xpos,\ypos)[#4`#5`{#7}]{\height}{\arrowtypea}a}%
\putmorphism(\xpos,\ypos)(1,-1)[`#6`{#8}]{\height}{\arrowtypeb}l%
\advance\xpos by\height
\advance\xpos by\height
\putmorphism(\xpos,\ypos)(-1,-1)[``{#9}]{\height}{\arrowtypec}r%
}}

\def\putVtriangle{\@ifnextchar <{\putVtrianglep}{\putVtrianglep
   <\arrowtypea`\arrowtypeb`\arrowtypec;\height>}}
\def\Vtriangle{\@ifnextchar <{\Vtrianglep}{\Vtrianglep
   <\arrowtypea`\arrowtypeb`\arrowtypec;\height>}}
 
\def\Vtrianglep<#1>[#2`#3`#4;#5`#6`#7]{{
\settriparms[#1]
\width=\height                         
\xext=\width                           
\yext=\height                          
\topadjust[#2`#3`{#5}]
\botadjust[#4``]
\multiply \xext by2 
\leftsladjust[#2`#3`{#6}]
\rightsladjust[#3`#4`{#7}]
\begin{picture}(\xext,\yext)(\xoff,\yoff)%
\putVtrianglep<\arrowtypea`\arrowtypeb`\arrowtypec;\height>%
(0,0)[#2`#3`#4;#5`#6`{#7}]%
\end{picture}%
}}
 
\def\putVtrianglepairp<#1>(#2)[#3;#4`#5`#6`#7`#8]{{
\settripairparms[#1]%
\setpos(#2)%
\settokens[#3]%
\advance\ypos by\height
\putmorphism(\xpos,\ypos)(1,-1)[`\tokend`{#6}]{\height}{\arrowtypec}l%
\puthmorphism(\xpos,\ypos)[\tokena`\tokenb`{#4}]{\height}{\arrowtypea}a%
\advance\xpos by\height
\putvmorphism(\xpos,\ypos)[``{#7}]{\height}{\arrowtyped}m%
\advance\xpos by\height
\putmorphism(\xpos,\ypos)(-1,-1)[``{#8}]{\height}{\arrowtypee}r%
}}
 
\def\putVtrianglepair{\@ifnextchar <{\putVtrianglepairp}{\putVtrianglepairp%
    <\arrowtypea`\arrowtypeb`\arrowtypec`\arrowtyped`\arrowtypee;\height>}}
\def\Vtrianglepair{\@ifnextchar <{\Vtrianglepairp}{\Vtrianglepairp%
    <\arrowtypea`\arrowtypeb`\arrowtypec`\arrowtyped`\arrowtypee;\height>}}
 
\def\Vtrianglepairp<#1>[#2;#3`#4`#5`#6`#7]{{%
\settripairparms[#1]%
\settokens[#2]
\xext=\height                  
\width=\height                 
\yext=\height                  
\vertadjust[\tokena`\tokenb`{#4}]
\tempcountd=\tempcounta        
\vertadjust[\tokenb`\tokenc`{#5}]
\ifnum\tempcounta<\tempcountd%
\tempcounta=\tempcountd\fi
\advance \yext by\tempcounta
\botadjust[\tokend``]%
\multiply \xext by2
\leftsladjust[\tokena`\tokend`{#6}]%
\rightsladjust[\tokenc`\tokend`{#7}]%
\begin{picture}(\xext,\yext)(\xoff,\yoff)%
\putVtrianglepairp
<\arrowtypea`\arrowtypeb`\arrowtypec`\arrowtyped`\arrowtypee;\height>%
(0,0)[#2;#3`#4`#5`#6`{#7}]%
\end{picture}%
}}

\def\putCtrianglep<#1>(#2,#3)[#4`#5`#6;#7`#8`#9]{{%
\settriparms[#1]%
\xpos=#2 \ypos=#3
\advance\ypos by\height
\putmorphism(\xpos,\ypos)(1,-1)[``{#9}]{\height}{\arrowtypec}l%
\advance\xpos by\height
\advance\ypos by\height
\putmorphism(\xpos,\ypos)(-1,-1)[#4`#5`{#7}]{\height}{\arrowtypea}l%
{\multiply\height by 2
\putvmorphism(\xpos,\ypos)[`#6`{#8}]{\height}{\arrowtypeb}r}%
}}
 
\def\putCtriangle{\@ifnextchar <{\putCtrianglep}{\putCtrianglep
    <\arrowtypea`\arrowtypeb`\arrowtypec;\height>}}
\def\Ctriangle{\@ifnextchar <{\Ctrianglep}{\Ctrianglep
    <\arrowtypea`\arrowtypeb`\arrowtypec;\height>}}
 
\def\Ctrianglep<#1>[#2`#3`#4;#5`#6`#7]{{
\settriparms[#1]
\width=\height                          
\xext=\width                            
\yext=\height                           
\multiply \yext by2 
\topadjust[#2``]
\botadjust[#4``]
\sladjust[#3`#2`{#5}]{\width}
\tempcountd=\tempcounta                 
\sladjust[#3`#4`{#7}]{\width}
\ifnum \tempcounta<\tempcountd          
\tempcounta=\tempcountd\fi              
\advance \xext by\tempcounta            
\advance \xoff by-\tempcounta           %
\rightadjust[#2`#4`{#6}]
\begin{picture}(\xext,\yext)(\xoff,\yoff)%
\putCtrianglep<\arrowtypea`\arrowtypeb`\arrowtypec;\height>%
(0,0)[#2`#3`#4;#5`#6`{#7}]%
\end{picture}%
}}
 
\def\putDtrianglep<#1>(#2,#3)[#4`#5`#6;#7`#8`#9]{{%
\settriparms[#1]%
\xpos=#2 \ypos=#3
\advance\xpos by\height \advance\ypos by\height
\putmorphism(\xpos,\ypos)(-1,-1)[``{#9}]{\height}{\arrowtypec}r%
\advance\xpos by-\height \advance\ypos by\height
\putmorphism(\xpos,\ypos)(1,-1)[`#5`{#8}]{\height}{\arrowtypeb}r%
{\multiply\height by 2
\putvmorphism(\xpos,\ypos)[#4`#6`{#7}]{\height}{\arrowtypea}l}%
}}
 
\def\putDtriangle{\@ifnextchar <{\putDtrianglep}{\putDtrianglep
    <\arrowtypea`\arrowtypeb`\arrowtypec;\height>}}
\def\Dtriangle{\@ifnextchar <{\Dtrianglep}{\Dtrianglep
   <\arrowtypea`\arrowtypeb`\arrowtypec;\height>}}
 
\def\Dtrianglep<#1>[#2`#3`#4;#5`#6`#7]{{
\settriparms[#1]
\width=\height                         
\xext=\height                          
\yext=\height                          
\multiply \yext by2 
\topadjust[#2``]
\botadjust[#4``]
\leftadjust[#2`#4`{#5}]
\sladjust[#3`#2`{#5}]{\height}
\tempcountd=\tempcountd                
\sladjust[#3`#4`{#7}]{\height}
\ifnum \tempcounta<\tempcountd         
\tempcounta=\tempcountd\fi             
\advance \xext by\tempcounta           %
\begin{picture}(\xext,\yext)(\xoff,\yoff)
\putDtrianglep<\arrowtypea`\arrowtypeb`\arrowtypec;\height>%
(0,0)[#2`#3`#4;#5`#6`{#7}]%
\end{picture}%
}}
 
\def\setrecparms[#1`#2]{\width=#1 \height=#2}%
\def\recursep<#1`#2>[#3;#4`#5`#6`#7`#8]{{%
\width=#1 \height=#2
\settokens[#3]
\settowidth{\tempdimen}{$\tokena$}
\ifdim\tempdimen=0pt
  \savebox{\tempboxa}{\hbox{$\tokenb$}}%
  \savebox{\tempboxb}{\hbox{$\tokend$}}%
  \savebox{\tempboxc}{\hbox{$#6$}}%
\else
  \savebox{\tempboxa}{\hbox{$\hbox{$\tokena$}\times\hbox{$\tokenb$}$}}%
  \savebox{\tempboxb}{\hbox{$\hbox{$\tokena$}\times\hbox{$\tokend$}$}}%
  \savebox{\tempboxc}{\hbox{$\hbox{$\tokena$}\times\hbox{$#6$}$}}%
\fi
\ypos=\height
\divide\ypos by 2
\xpos=\ypos
\advance\xpos by \width
\xext=\xpos \yext=\height
\topadjust[#3`\usebox{\tempboxa}`{#4}]%
\botadjust[#5`\usebox{\tempboxb}`{#8}]%
\sladjust[\tokenc`\tokenb`{#5}]{\ypos}%
\tempcountd=\tempcounta
\sladjust[\tokenc`\tokend`{#5}]{\ypos}%
\ifnum \tempcounta<\tempcountd
\tempcounta=\tempcountd\fi
\advance \xext by\tempcounta
\advance \xoff by-\tempcounta
\rightadjust[\usebox{\tempboxa}`\usebox{\tempboxb}`\usebox{\tempboxc}]%
\bfig
\putCtrianglep<-1`1`1;\ypos>(0,0)[`\tokenc`;#5`#6`{#7}]%
\puthmorphism(\ypos,0)[\tokend`\usebox{\tempboxb}`{#8}]{\width}{-1}b%
\puthmorphism(\ypos,\height)[\tokenb`\usebox{\tempboxa}`{#4}]{\width}{-1}a%
\advance\ypos by \width
\putvmorphism(\ypos,\height)[``\usebox{\tempboxc}]{\height}1r%
\efig
}}
 
\def\recurse{\@ifnextchar <{\recursep}{\recursep<\width`\height>}}
 
\def\puttwohmorphisms(#1,#2)[#3`#4;#5`#6]#7#8#9{{%
\puthmorphism(#1,#2)[#3`#4`]{#7}0a
\ypos=#2
\advance\ypos by 20
\puthmorphism(#1,\ypos)[\phantom{#3}`\phantom{#4}`#5]{#7}{#8}a
\advance\ypos by -40
\puthmorphism(#1,\ypos)[\phantom{#3}`\phantom{#4}`#6]{#7}{#9}b
}}
 
\def\puttwovmorphisms(#1,#2)[#3`#4;#5`#6]#7#8#9{{%
\putvmorphism(#1,#2)[#3`#4`]{#7}0a
\xpos=#1
\advance\xpos by -20
\putvmorphism(\xpos,#2)[\phantom{#3}`\phantom{#4}`#5]{#7}{#8}l
\advance\xpos by 40
\putvmorphism(\xpos,#2)[\phantom{#3}`\phantom{#4}`#6]{#7}{#9}r
}}
 
\def\puthcoequalizer(#1)[#2`#3`#4;#5`#6`#7]#8#9{{%
\setpos(#1)%
\puttwohmorphisms(\xpos,\ypos)[#2`#3;#5`#6]{#8}11%
\advance\xpos by #8
\puthmorphism(\xpos,\ypos)[\phantom{#3}`#4`#7]{#8}1{#9}
}}
 
\def\putvcoequalizer(#1)[#2`#3`#4;#5`#6`#7]#8#9{{%
\setpos(#1)%
\puttwovmorphisms(\xpos,\ypos)[#2`#3;#5`#6]{#8}11%
\advance\ypos by -#8
\putvmorphism(\xpos,\ypos)[\phantom{#3}`#4`#7]{#8}1{#9}
}}
 
\def\putthreehmorphisms(#1)[#2`#3;#4`#5`#6]#7(#8)#9{{%
\setpos(#1) \settypes(#8)
\if a#9 %
     \vertsize{\tempcounta}{#5}%
     \vertsize{\tempcountb}{#6}%
     \ifnum \tempcounta<\tempcountb \tempcounta=\tempcountb \fi
\else
     \vertsize{\tempcounta}{#4}%
     \vertsize{\tempcountb}{#5}%
     \ifnum \tempcounta<\tempcountb \tempcounta=\tempcountb \fi
\fi
\advance \tempcounta by 60
\puthmorphism(\xpos,\ypos)[#2`#3`#5]{#7}{\arrowtypeb}{#9}
\advance\ypos by \tempcounta
\puthmorphism(\xpos,\ypos)[\phantom{#2}`\phantom{#3}`#4]{#7}{\arrowtypea}{#9}
\advance\ypos by -\tempcounta \advance\ypos by -\tempcounta
\puthmorphism(\xpos,\ypos)[\phantom{#2}`\phantom{#3}`#6]{#7}{\arrowtypec}{#9}
}}
 
\def\putarc(#1,#2)[#3`#4`#5]#6#7#8{{%
\xpos #1
\ypos #2
\width #6
\arrowlength #6
\putbox(\xpos,\ypos){#3\vphantom{#4}}%
{\advance \xpos by\arrowlength
\putbox(\xpos,\ypos){\vphantom{#3}#4}}%
\horsize{\tempcounta}{#3}%
\horsize{\tempcountb}{#4}%
\divide \tempcounta by2
\divide \tempcountb by2
\advance \tempcounta by30
\advance \tempcountb by30
\advance \xpos by\tempcounta
\advance \arrowlength by-\tempcounta
\advance \arrowlength by-\tempcountb
\halflength=\arrowlength \divide\halflength by 2
\divide\arrowlength by 5
\put(\xpos,\ypos){\bezier{\arrowlength}(0,0)(50,50)(\halflength,50)}
\ifnum #7=-1 \put(\xpos,\ypos){\vector(-3,-2)0} \fi
\advance\xpos by \halflength
\put(\xpos,\ypos){\xpos=\halflength \advance\xpos by -50
   \bezier{\arrowlength}(0,50)(\xpos,50)(\halflength,0)}
\ifnum #7=1 {\advance \xpos by
   \halflength \put(\xpos,\ypos){\vector(3,-2)0}} \fi
\advance\ypos by 50
\vertsize{\tempcounta}{#5}%
\divide\tempcounta by2
\advance \tempcounta by20
\if a#8 %
   \advance \ypos by\tempcounta
   \putbox(\xpos,\ypos){#5}%
\else
   \advance \ypos by-\tempcounta
   \putbox(\xpos,\ypos){#5}%
\fi
}}
 
\makeatother

\title{First-Order Modal Logic via Logical Categories
}
\author{Silvio~Ghilardi and J\'er\'emie~Marqu\`es}
\address{Department of Mathematics, Universit\`a degli Studi di Milano, Italy}
\date{\today}

\begin{document}

\begin{abstract}
We extend the logical categories framework to first order modal logic. In our \emph{modal categories},
 modal operators are applied directly to subobjects and interact with the background factorization system. We prove a Joyal-style representation theorem into relational structures formalizing a `counterpart' notion. We investigate saturation conditions related to definability questions 
 and we enrich our framework with quotients and disjoint sums, thus leading to the notion of a \emph{modal (quasi) pretopos}.
 We finally show how to build syntactic categories out of first order modal theories.
\end{abstract}


\maketitle

\section{Introduction}
\label{sec:intro}

Modal logic is a pervasive formalism with applications in computer science, mathematics and philosophy. The core modal system is the minimal normal modal logic system (called $K$) which is axiomatized at the propositional level by adding to classical tautologies and modus ponens the distribution law
\begin{equation}\label{eq:K1}
 \Box (\phi \to \psi) \to (\Box \phi \to \Box \psi)
\end{equation}
and the necessitation rule
\begin{equation}\label{eq:K2}
{\rm from}~~~\phi~~~~~{\rm infer}~~~~~\Box \phi~~.
\end{equation}
Algebraically, this axiomatization corresponds (via standard Lindembaum constructions) to \emph{modal algebras} which are  Boolean algebras endowed with a unary operator subject to the equations
$$
\Box (x\wedge y) = \Box x \wedge \Box y \qquad \Box \top =\top~~.
$$
Using equivalently the dual `possibility' operator $\Diamond := \neg \Box \neg$ such equations can be rewritten as
$$
\Diamond (x\vee y) = \Diamond x \vee \Diamond y \qquad \Diamond \bot =\bot~~.
$$
In the applications, basic modal operators satisfying the axioms of $K$ serve as building blocks for more complex modalities obtained by combining them with each other and by applying various kinds of constructors, typically fixpoint constructors.
In this way, many logics like PDL~\cite{pdl}, LTL, CTL, CTL$^*$~\cite{temporal_logic} (culminating in the $\mu$-calculus~\cite{mu_calculus}) arose in the formal verification area. Similarly, in the knowledge representation area, basic $K$-operators are combined to give rise to various description logics like $\mathcal{ALC}$ and extensions~\cite{description_logics}.

In mathematical frameworks, like topology, the typical modal system is the system $S4$ obtained by adding to $K$ the further `reflexivity' and `transitivity' axioms
$$
\Box \phi \to \phi, \qquad \Box \phi \to \Box \Box \phi~~;
$$
such axioms correspond, at the algebraic level, to  \emph{interior} algebras, i.e. to modal algebras whose $\Box$ operator satisfies precisely the axioms of the interior operator in topological spaces~\cite{tarski}.

Most of the literature in modal logic is confined to propositional systems. However, predicate extensions turn our to be important too, e.g.: (i) in the formal verification area, where relative completeness~\cite{manna} brings together temporal and theory reasoning; (ii) in the knowledge representation area, where decidability  results can surprisingly be obtained for some expressive fragments of predicate modal logics, like the monadic fragments~\cite{many_dimensional,handbook}.

However, a categorical approach via logical categories\footnote{By `logical categories' we mean regular categories, coherent categories, pretoposes, etc. see~\cite{MR}.}  to quantified modal logic has to face the structural problem of the \emph{lack of homogeneity between interpretation of sorts and predicates} that arises in standard semantics like Kripke semantics. The problem can be explained as follows. A predicate Kripke frame~\cite{Val} is a triple  $(W,R, D)$, where $W$ is a set, $R\subseteq W\times W$ is a relation and $D$ is a function associating a set $D_w$ with every $w\in W$, with the condition $W_w\subseteq W_v$ in case $wRv$. In a predicate Kripke frame, an $n$-ary relation symbol $P$ is interpreted as a collection of subsets $P_w\subseteq D_w^n$ indexed by $w\in W$. Thus, contrary to what happens for intuitionistic logic, the interpretation of a relation symbol \emph{is not a predicate Kripke frame itself} and this is incompatible with the categorical logic practice  of interpreting relation symbols (and more generally formulae) as subobjects.

A possible way out would be the adoption of Lawvere doctrines~\cite{hyperdoctrines} as a categorical framework: a modal Lawvere doctrine could be easily defined 
(see e.g.~\cite{shirasu})
as a pair given by a category with products and a contravariant functor with values in the category of modal algebras (suitable conditions like the Beck-Chevalley condition should also be added). However, this solution (if not accompanied by suitable comprehension principles)
would just certify  the above lack of homogeneity between the interpretation of terms and formulae; in addition, it would  contradict one of the main principles of the categorical approach to logic, namely the presentation independence. After all, whether a certain definable set is logically handled using a sort symbol or a  formula is a question of presentations (analogously, whether a definable function is mirrored by a term or by a functional relation is also a matter of presentation). Unfortunately, it is precisely comprehension which fails in the above mentioned semantic modal contexts.

A more intriguing solution was proposed by F.W. Lawvere and implemented in~\cite{reyes}. The asymmetry between sorts and predicates is conceptually explained by saying  that our semantics involves in fact \emph{two} toposes and a geometric morphism connecting them: from one of the two toposes, one takes the interpretation of sorts and function symbols and from the other one, one takes the interpretation of relation symbols and formulae. This schema encompasses as a special case Kripke semantics, its generalization to presheaves~\cite{louvain,incompleteness} and also the interpretation of the modality as interior in \'etale spaces~\cite{handbook,awodey_rsl,awodey_aiml}. A representation theorem in Joyal style~\cite{MR} for pairs of logical categories connected by a functor having local adjoints was proved in~\cite{makkai_reyes_apal}.

This solution has the major merit of not introducing modalities as extra features, but of generating them out of the pure basic category-theoretic machinery, as it happens with all the classical and intuitionistic logical operators.  Still, the solution suffers from a considerable drawback: \emph{$S4$ axioms are forced} whenever modalities are produced by applying the above schema involving two different categories connected by a functor. This is unsatisfactory because, as pointed out above, the weaker system $K$ seems to be involved in most applications outside the realm of pure mathematics. However, precisely turning back to a pure mathematical context,  another way out appears. In topological models, modalities (namely the interior or its dual closure operator) apply to subsets, aka subspaces. Subspaces are \emph{regular subobjects}, i.e. subsets with the induced topology. This simple observation reveals an unexpected solution, whose main feature is that of \emph{changing the underlying factorization system} from the regular epi/mono factorization system adopted in logical categories, to an abstract factorization system, satisfying minimal conditions (like pullback stability).

One may reach the same conclusion by generalizing Kripke semantics too.
A long-debated question among philosophers~\cite{lewis,hazen} is the trans-world \emph{re-identification} problem. In Kripke semantics, the question is handled trivially: it is assumed that $D_w\subseteq D_v$ whenever we have $wRv$. This means that
some $a\in D_w$ is identified with itself in all the accessible worlds $v$. However, this is no longer true if we generalize predicate Kripke frames to presheaves, because there the transition from $D_w$ to $D_v$ takes the form of a function or even of a multiplicity of functions, if presheaves are taken over a category which is not a poset. Once the Kripkean inclusion trivialization is broken, nothing prevents from taking \emph{relations} as transitions from a domain $D_v$ to a domain $D_w$: such relations are called `counterpart relations.'
In the resulting semantics, the forcing condition for $\Diamond$  becomes ``$a\in D_w$ satisfies $\Diamond P$ iff some counterpart $b\in D_v$ of $a$ (with $w R v$) satisfies $P$'' (see the defining condition~\eqref{eq:counterpart} in Section~\ref{sec:modalities} below).
A formal realization of this semantics was introduced in~\cite{louvain}; the universes generalizing predicate Kripke frames and presheaves were there called `relational presheaves' and further analyzed e.g. in
~\cite{handbook,Niefield,Niefield1,Gadducci}. What we want to point out now  is that  the asymmetry between interpretation of sorts and of predicates disappears in such relational presheaves, because  a `collections of subsets indexed by possible worlds' is, once again, just  a regular mono (this is because relations, unlike inclusions or functions, can be restricted to subsets).

In Section~\ref{sec:modalities} we take up relational presheaves as our fundamental semantic environment; however we shall speak of `relational $\bf G$-sets' instead, because we assume the variation domain to be a graph $\bf G$ and not a category. This choice was already suggested in~\cite{louvain}  and is needed in order to avoid the validity of $S4$ axioms; moreover, we shall see that we get 
the surprisingly much richer categorical structure of a \emph{quasi-topos}
(see Proposition~\ref{prop:quasitopos} below) by adopting relational $\bf G$-sets instead of relational presheaves, which motivates once more  the
relevance of   $K$ modalities.

The paper is structured as follows: in Section~\ref{sec:F} we review some preliminary material on factorization systems, in Section~\ref{sec:modalities} we introduce modal categories and supply examples of them.  Then we prove a representation theorem in Joyal style for modal categories into relational $\bf G$-sets (Section~\ref{sec:representation}).
 The paper continues with a detailed analysis of saturation aspects in modal categories~\ref{sec:comprehension} and 
 then it extends the representation theorem to
 quotients and disjoint unions, i.e. to
 modal quasi-pretoposes (Section~\ref{sec:quotients}).
Lastly, in Section~\ref{sec:theories} we discuss the construction of syntactic categories out of modal first order theories. We conclude the paper with a discussion on further developments (Section~\ref{sec:conclusions}).

\textbf{Notations.} In this paper, the composition of two arrows $f : X\lra Y$ and $g : Y\lra Z$ is written $fg$ or $g\circ f$. When we speak of semilattices, we always mean meet-semilattices with unit and when we speak of lattices we always mean lattices with unit and zero.

\section{Background on Factorization Systems}\label{sec:F}

Coherent categories and pretoposes have been thoroughly studied as the syntactic categories of first-order coherent theories. Several logics have been given a similar categorical presentation. Among the most satisfying examples, we can mention geometric logic with Grothendieck toposes, higher-order intuitionistic logic with elementary toposes, and intuitionistic logic with Heyting pretoposes. 

We will define here \emph{modal categories} in a similar spirit, as categories in which first-order modal logic can be interpreted. 
When $X$ is an object of a coherent category, the `propositions' in context $X$ are all the monomorphisms with codomain $X$. For modal logic, however, we need to add some extra structure. First, we need to specify which monomorphisms should be interpreted as `propositions': for instance a `proposition' on a topological space should be a subset of the underlying set, hence a subspace, and not all the monomorphisms are subspace embeddings. Second, we need to specify how the modalities act on these propositions/subspaces.

In order to account for the fact that not all the monomorphisms are treated as `subspaces,' we need to relax the usual categorical notions of regular and coherent categories. Yet, we will maintain the important property that every morphism factors as a `surjection' followed by a `subspace embedding.' Hence, we start with  orthogonal factorization systems. We recall some facts about them below, but we refer the reader to \cite[\S~3.1]{GhilardiZawadowski2011} and \cite[\S~5.5]{BorHandbookCategoricalAlgebra1994} for more details.

Let $\cE$ be a lex category. An orthogonal factorization system on $\cE$ can be defined as a pair $(\Ec,\Mc)$ of classes of morphisms of $\cE$ such that:
\begin{itemize}
	\item Each morphism of $\cE$ can be written as $em$ with $e \in \Ec$ and $m \in \Mc$.
	\item Both $\Ec$ and $\Mc$ contain identities and are closed under left and right composition with isomorphisms.
	\item Every arrow of $\Ec$ is orthogonal on the left to every arrow of $\Mc$. In other words, for each commutative square
	\[\begin{tikzcd}
		A \ar[r,"e"] \ar[d,"u"'] & B \ar[d,"v"]\\
		C \ar[r,"m"] & D
	\end{tikzcd}\]
	where $e \in \Ec$ and $m \in \Mc$, there is a unique diagonal filler $w : B\superto C$ such that $ew = u$ and $wm = v$.
\end{itemize}
In an orthogonal factorization system, the factorization of a morphism as $e\,m$ with $e \in \Ec$ and $m \in \Mc$ is unique up to a unique isomorphism, and both $\Ec$ and $\Mc$ are stable under composition.

In order to interpret first-order regular logic in $\cE$, we need to impose some further conditions. The most essential one is stability: An orthogonal factorization system $(\Ec,\Mc)$ is called \emph{stable} if $\Ec$ and $\Mc$ are stable under pullbacks. (In fact, $\Mc$ is necessarily stable under pullbacks in an orthogonal factorization system.)

\begin{definition}\label{dfn:f-reg-cat}
	An \emph{f-regular category} (short for `factorization-regular category') is a lex category $\cE$ equipped with a stable orthogonal factorization system $(\Ec,\Mc)$ such that $\Mc$ is included in the class of all monomorphisms, and contains the class of regular monomorphisms. A lex functor between f-regular categories is \emph{f-regular} when it preserves the factorization system.
\end{definition}

In Definition~\ref{dfn:f-reg-cat}, the condition that all the arrows in $\Mc$ are monomorphic ensures that we can think of them as subobjects. As we will see in \S~\ref{subsec:hyperdoc}, this allows us to interpret regular first-order logic in $\cE$ by means of a Lawvere doctrine. In these terms, the additional requirement that every regular monomorphism is in $\Mc$ means that if two parallel arrows $f$ and $g$ satisfy $\vdash f(x)=g(x)$ in the internal language, then $f=g$.

Since $\Mc$ suffices to determine $\Ec$ 
(because $\Ec$ turns out to be the left orthogonal class $\Mc^\bot$ to $\Mc$), we will denote in general an f-regular category by the pair $(\cE,\Mc)$, sometimes even leaving $\Mc$ implicit. Every regular category is f-regular, with the factorization system of regular epimorphisms and monomorphisms. Every r-regular category, as defined in \cite[\S~3.1]{GhilardiZawadowski2011}, is an f-regular category with the factorization system of epimorphisms and regular monomorphisms. For instance, the category $\bf Top$ of topological spaces and continuous function is r-regular.

The followng proposition shows that in an \Fregular category, the left class $\Ec$ satisfies the dual properties of the class $\cM$:

\begin{proposition}\label{prop:epis}
    In an \Fregular category with factorization system $(\Ec, \cM)$, we have that $\Ec$ includes regular epis and is included in the class of all epis.
\end{proposition}

\begin{proof}
    Let  $p$ be the coequalizer of $g_1,g_2$ and let $m\circ q$ be a $(\Ec, \cM)$-facto\-ri\-za\-tion of $p$. By the universal property of coequalizers there is $h$ such that $h\circ p=q$. Then $m\circ h\circ p=m\circ q= p$ and so $m\circ h=\id$ because $p$ is epi being a coequalizer. Now $m$ is iso because it is mono (recall $m\in \cM$) and split epi. This shows that $p\in \Ec$.

    Let now $p\in \Ec$ be such that $g_1\circ p = g_2\circ p$ and take the equalizer $m$ of $g_1,g_2$. Then there is
    $h$ such that $m\circ h=p$. Since $m\in \cM$ and $m$ is orthogonal to $p$ there is $l$ such that $m\circ l=id$.
    But then $m$ is iso (being split epi and mono), which means that $g_1=g_2$.
\end{proof}

\subsection{The Lawvere doctrine of $\Mc$-subobjects}
\label{subsec:hyperdoc}

In an f-regular category $(\cE,\Mc)$, the arrows in $\Mc$ with codomain $X$ (better, their equivalence
classes under isomorphisms) are called the \emph{$\Mc$-subobjects} of $X$ and we denote them as 
$S\buildrel s\over\hookrightarrow X$. We think of $S$ as a `subspace' of $X$. The $\Mc$-subobjects of $X$ 
are preordered in the standard way: Given $s_1 : S_1 \hookrightarrow X$ and $s_2 : S_2 \hookrightarrow 
X$, we write $S_1 \leq S_2$ if there is $h : S_1 \superto S_2$ such that $hs_2 = s_1$. Since $s_2$ is 
monomorphic, such an $h$ is unique, and it is also in $\Mc$ by Lemma~\ref{lem:firstcomponent} below.

\begin{lemma}\label{lem:firstcomponent}
In an \Fregular category $(\cE,\cM)$, the first component of a map in $\cM$ is in $\cM$ too.
\end{lemma}

\begin{proof}
        Let $hl\in \Mc$ and
	write $h$ as $em$ with $e \in \Mc^\bot$ and $m \in \Mc$. By orthogonality of $e$ and $hl$, we obtain an arrow $f$ making the following diagram commute.
	\[\begin{tikzcd}
		\cdot \ar[r,"e"] \ar[d,"\id"'] & \cdot \ar[d,"ml"] \ar[ld,dashed,"f"']\\
		\cdot \ar[r,"hl"'] & \cdot
	\end{tikzcd}\]
	This means that $e$ is a split monomorphism, hence it is in $\Mc$, hence it is an isomorphism because it is also in $\Mc^\bot$. Thus $h = em \in \Mc$.
\end{proof}

If $S_1 \leq S_2$ and $S_2 \leq S_1$, then $S_1$ and $S_2$ differ by an isomorphism and we say that they are equivalent as $\Mc$-subobjects: we denote by $\Sub_\Mc(X)$ the equivalence classes of $\Mc$-subobjects of $X$. Actually, by pullback stability, taking pullbacks yields a contravariant functor to the category of meet-semilattices. We indicate with $f^* : \Sub_\Mc(Y) \superto \Sub_\Mc(X)$ the operation of pulling back along $f : X\superto Y$. More is true indeed: we get a regular Lawvere doctrine structure, as defined below.


\begin{definition}\label{dfn:BC-sq}
	Let
	\[\begin{tikzcd}
		A \ar[r,"f^*"] \ar[d,"g^*"'] & B \ar[d,"u^*"]\\
		C \ar[r,"v^*"] & D
	\end{tikzcd}\]
	be a commutative square in the category of posets. Suppose that $g^*$ and $u^*$ have left adjoints $\exists_g$ and $\exists_u$. We say that the square is \emph{Beck--Chevalley} if $f^*\circ\exists_g = \exists_u\circ v^*$.
\end{definition}


Below, we let $\cSMLat$ be the category of semilattices and related morphisms.

\begin{definition}
	A \emph{regular  Lawvere doctrine} 
    is a pair $(\cE, \Pc)$ given by a 
     lex category $\cE$ and a functor $\Pc : {\cE}^{op}\superto\cSMLat$ such that:
	\begin{itemize}
		\item For every $f : X\superto Y$ in $\cE$, the function $f^* = \Pc(f) : \Pc(Y)\superto\Pc(X)$ has a left adjoint $\exists_f : \Pc(X) \superto \Pc(Y)$.
		\item For every $f : X\superto Y$ in $\cE$, the so-called Frobenius condition holds:
		\[ \exists_f(\varphi \land  f^*(\psi)) = (\exists_f \varphi) \land \psi \text{.} \]
		\item Every pullback square is sent to a Beck--Chevalley square.
	\end{itemize}
\end{definition}

\begin{proposition}\label{p31.1}
	If $(\cE,\Mc)$ is an f-regular category, then $\Sub_\Mc$ is a regular Lawvere doctrine.
\end{proposition}

\begin{proof}
	Let $f : X\superto Y$ be an arrow in $\cE$. We first show that $f^* : \Sub_\Mc(Y) \superto \Sub_\Mc(X)$ has a left adjoint $\exists_f : \Sub_\Mc(X) \superto \Sub_\Mc(Y)$. Let $s : S \hookrightarrow X$ in $\Mc$. Then $\exists_f S \in \Sub_\Mc(Y)$ is obtained by factoring $sf$ as an arrow $S \superto \exists_f S$ in $\Ec$ followed by an arrow $\exists_f S \hookrightarrow Y$ in $\Mc$. Let $T \hookrightarrow Y$ in $\Mc$. We want to show that in the diagram below, the dashed arrow $u$ exists if and only if the dashed arrow $v$ exists.
	\[\begin{tikzcd}
		S \ar[dd,hook] \ar[rr] \ar[rd,dashed,"u"] & & \exists_f S \ar[dd,hook] \ar[rd,dashed,"v"] &\\
		& f^* T \ar[rr,crossing over] \ar[ld,hook] & & T \ar[ld,hook] \\
		X \ar[rr,"f"'] & & Y &
	\end{tikzcd}\]
	If $v$ exists, then $u$ exists too by the universal property of the pullback defining $f^* T$. On the other hand, if $u$ exists, then $v$ exists too by applying the orthogonality of $S \superto \exists_f S$ and $T \hookrightarrow Y$ in the diagram below.
	\[\begin{tikzcd}
		S \ar[r] \ar[d] & \exists_f S \ar[dd,hook]\\[-1em]
		f^* T \ar[d] & \\[-1em]
		T \ar[r,hook] & Y
	\end{tikzcd}\]
	
	In order to check the Beck--Chevalley condition, consider a commutative square as below in $\cE$.
	\[\begin{tikzcd}
		X \ar[r,"p_2"] \ar[d,"p_1"'] & Y_2 \ar[d,"f_2"]\\
		Y_1 \ar[r,"f_1"] & Z
	\end{tikzcd}\]
	Given $s : S \hookrightarrow Y_1$ in $\Mc$, take the further pullback below.
	\[\begin{tikzcd}
		p^*_1 S \ar[r,"s'",hook] \ar[d] & X \ar[d,"p_1"] \\
		S \ar[r,hook,"s"] & Y_1
	\end{tikzcd}\]
	We have to show that the $(\Ec,\Mc)$-factorization of $s' p_2$ is just (up to an 	isomorphism) the factorization of $s f_1$ pulled back along $f_2$. But this is just the stability the factorization applied to the pullback square below.
	\[\begin{tikzcd}
		p^*_1 S \ar[r,"s' p_2"] \ar[d] & Y_2 \ar[d,"f_2"]\\
		S \ar[r,"s f_1"] & Z
	\end{tikzcd}\]
	
	As for the Frobenius condition, it is actually a special case of the Beck--Chevalley condition. Let $f : X\superto Y$ in $\cE$. Given $S \in \Sub_\Mc(Y)$, we form the pullback square below.
	\[\begin{tikzcd}
		f^* S \ar[r] \ar[d,hook] & S \ar[d,hook]\\
		X \ar[r,"f"] & Y
	\end{tikzcd}\]
	We then combine the Beck--Chevalley condition applied to this square and the two facts:
	\begin{itemize}
		\item pulling back an arrow of $\Mc$ along another arrow of $\Mc$ is an intersection in $\Sub_\Mc(X)$;
		\item if $S\hookrightarrow Y$ is in $\Mc$, the factorization of an arrow $Z\superto S$ is obtained by factoring $Z\superto S\hookrightarrow Y$ and restricting to $S$.
	\end{itemize}
	We obtain, for all $T \in \Sub_\Mc(X)$, the Frobenius condition
	\[ \exists_f (T \land f^*S) = (\exists_f T) \land S \text{.} \]
\end{proof}


In fact, as shown in Proposition~\ref{prop:f-reg-sub-hyperdoc} below, f-regular categories form in this way a (non-full) sub-category of regular Lawvere doctrines. Given two regular Lawvere doctrines $\Pc_1 : {\cE}_1^{op}\superto\cSMLat$ and $\Pc_2 : {\cE}_2^{op}\superto\cSMLat$, a \emph{morphism} $\Pc_1 \superto \Pc_2$ is a pair $(\Phi, \alpha)$ given by  lex functor $\Phi : {\cE}_1 \superto {\cE}_2$ and a natural transformation $\alpha : \Pc_1 \superto \Pc_2 \circ \Phi$ such that all the naturality squares below (induced by some arrow $X\lra Y$ in ${\cE}_1$)
are Beck--Chevalley.
\[\begin{tikzcd}
	\Pc_1(Y) \ar[r,"\alpha_m"] \ar[d] & \Pc_2(\Phi(Y)) \ar[d]\\
	\Pc_1(X) \ar[r,"\alpha_n"] & \Pc_2(\Phi(X))
\end{tikzcd}\]

\begin{proposition}\label{prop:f-reg-sub-hyperdoc}
	Let $({\cE}_1,\Mc_1)$ and $({\cE}_2,\Mc_2)$ be two f-regular categories. If $\Phi : {\cE}_1 \superto {\cE}_2$ is f-regular, then there is a unique natural transformation $\alpha : \Sub_{\Mc_1} \superto \Sub_{\Mc_2} \circ \Phi$ such that $(\Phi,\alpha)$ is a morphism of Lawvere doctrines.
\end{proposition}

\begin{proof}
	Let $\Phi : {\cE}_1 \superto {\cE}_2$ be an f-regular functor. We first show that there is at most one $\alpha : \Sub_{\Mc_1} \superto \Sub_{\Mc_2} \circ \Phi$ such that $(\Phi,\alpha)$ is a morphism of Lawvere doctrines. Let $S \in \Sub_{\Mc_1}(X)$. Consider the following naturality square of a natural transformation $\alpha : \Sub_{\Mc_1} \superto \Sub_{\Mc_2} \circ \Phi$.
	\[\begin{tikzcd}
		\Sub_{\Mc_1}(X) \ar[r,"\alpha_X"] \ar[d] & \Sub_{\Mc_2}(\Phi(X)) \ar[d]\\
		\Sub_{\Mc_1}(S) \ar[r,"\alpha_S"] & \Sub_{\Mc_2}(\Phi(S))
	\end{tikzcd}\]
	The Beck--Chevalley condition applied to $1_S \in \Sub_{\Mc_1}(S)$ says that $\alpha_X(S) = \Phi(S)$
    (i.e. $\alpha_X$ maps $S\hookrightarrow X$  to the subobject $\Phi(S) \hookrightarrow \Phi(X)$).
    Hence $\alpha$ is uniquely determined by $\Phi$ and $\alpha_X(S) = \Phi(S)$.
	
	Next, we check that putting $\alpha_X(S) = \Phi(S)$ always defines a natural transformation whose naturality squares are Beck--Chevalley. First, $\Phi(S) \in \Sub_{\Mc_2}(\Phi(X))$ whenever $S \in \Sub_{\Mc_1}(X)$ since $\Phi$ preserves $\Mc$. Since $\Phi$ is left exact, $\alpha_X$ preserves meets. Moreover, $\Phi$ preserves pullbacks, hence $\Phi(f)^* \Phi(S) = \Phi(f^* S)$ and $\alpha$ is a natural transformation. In order to check that the naturality squares are Beck--Chevalley, let $f : X\superto Y$ in ${\cE}_1$ and let $S \in \Sub_{\Mc_1}(X)$. Since $\Phi$ preserves the factorization of arrows, we obtain that $\Phi(\exists_f S) = \exists_{\Phi(f)} \Phi(S)$. This concludes the proof that $(\Phi,\alpha)$ is a morphism of Lawvere doctrines.
\end{proof}

We now show a converse of Proposition~\ref{prop:f-reg-sub-hyperdoc} when $({\cE}_2,\Mc_2)$ comes from a regular category such as $\bf Set$.

\begin{proposition}\label{prop:reg-full-hyperdoc}
	Let $({\cE}_1,\Mc_1)$ and $({\cE}_2,\Mc_2)$ be two f-regular categories and suppose that $\Mc_2$ is exactly the class of monomorphisms of ${\cE}_2$. Then every Lawvere doctrine morphism $\Sub_{\Mc_1} \superto \Sub_{\Mc_2}$ is induced by an f-regular functor ${\cE}_1 \superto {\cE}_2$ as in Proposition~\ref{prop:f-reg-sub-hyperdoc}.
\end{proposition}

\begin{proof}
	Let $\Phi : {\cE}_1 \superto {\cE}_2$ be a lex functor and let $\alpha : \Sub_{\Mc_1} \superto \Sub_{\Mc_2} \circ \Phi$ be a natural transformation whose naturality squares are Beck--Chevalley. We want to show that $\Phi$ respects the factorization systems. Any $f \in \Mc_1$ is a monomorphism, hence $\Phi(f)$ is also a monomorphism and it is in $\Mc_2$. Suppose now that $f : X\superto Y$ is in the left class of ${\cE}_1$. This means that $\exists_f X = Y$. Using that $\alpha$ preserves the top elements and applying the Beck--Chevalley condition on the naturality square associated to $f$, we obtain that $\exists_{\Phi(f)} \Phi(X) = \Phi(Y)$, which means that $\Phi(f)$ is in the left class of ${\cE}_2$.
\end{proof}

\begin{remark}
	The stable orthogonal factorization systems $(\Ec,\Mc)$ on a lex category $\cE$ such that $\Mc$ is included in the class of monomorphisms correspond to the regular Lawvere doctrines with \emph{full comprehension} as defined in 
    \cite{MaiRosQuotientCompletionFoundation2013,pasquali}. The class of regular monomorphisms is included in $\Mc$ if and only if every morphism of $\cE$ is determined uniquely by its graph in the internal language. 
\end{remark}


Coherent logic is obtained from regular logic by adding finite joins. For instance, a \emph{coherent Lawvere doctrine} is a regular doctrine valued in the category of distributive lattices. We define similarly f-coherent categories:

\begin{definition}
	An \emph{f-coherent category} is an f-regular category $(\cE,\Mc)$ whose functor $\Sub_{\Mc}$ is valued in the category of distributive lattices. An f-regular functor between f-coherent categories is \emph{f-coherent} when it moreover preserves finite joins of $\Mc$-subobjects. We say that $(\cE,\Mc)$ is \emph{f-Boolean} when $\Sub_\Mc$ is valued in the category of Boolean algebras.
\end{definition}

 \section{Adding Modalities}\label{sec:modalities}

A \emph{modal lattice} is a distributive lattice  $(D, \wedge, \top, \vee, \bot)$ endowed with an operator $\Diamond: D\longrightarrow D$ s.t.\ the equalities
$$
\Diamond(x\vee y)=\Diamond x \vee \Diamond y,\qquad \Diamond \bot=\bot
$$
hold for all $x,y\in D$. The typical example of a modal lattice (which is also a Boolean algebra) comes from a \emph{Kripke frame}, which is a pair $(W,R)$ given by a set endowed with a binary relation $R\subseteq W\times W$. Given such a Kripke frame, we can define a diamond operator:
$$
\Diamond_R: \wp(W) \longrightarrow \wp(W)
$$
by taking
\begin{equation}\label{eq:diamond}
\Diamond_R (S)~:=~\{w\in W\mid {\text{there is}}~ v\in S ~\text{s.t.}~ wRv\}~~.
\end{equation}
Notice that the definition of $\Diamond_R$ makes sense also whenever the domain and the codomain of the relation $R$ do not coincide: in such a case, if  
$R\subseteq W'\times W$, 
then the operator
\begin{equation}\label{eq:diamond1}
 \Diamond_R: \wp(W) \longrightarrow \wp(W')
\end{equation}
is a hemimorphism, i.e.\ it  preserves finite joins. In the following, we shall make use of operators like~\eqref{eq:diamond1} defined via~\eqref{eq:diamond} for relations
$R\subseteq W'\times W$.

This is our main definition:

\begin{definition}\label{def:main}
  A \emph{modal category} is an f-coherent category $(\cE,\Mc)$ such that for every object $X$, the lattice $\Sub_{\cM}(X)$ is a modal lattice. Moreover, the following conditions must be satisfied:
  \begin{itemize}
     \item \emph{(Continuity)} for every $f:X\longrightarrow Y$ and $S\in \Sub_{\cM}(Y)$ we have
      \begin{equation}\label{eq:continuity}
        \Diamond f^* S \leq f^* \Diamond S;
      \end{equation}
      \item \emph{(Subspace)} For every $m:A\hookrightarrow X$ in $\cM$ and for every
      $S\in \Sub_{\cM}(A)$ it holds that
      \begin{equation}\label{eq:subspace}
      \Diamond S=m^* \Diamond\, \exists_{m} \,S.
      \end{equation}
  \end{itemize}
\end{definition}

The two axioms above
are inspired by the definition of a continuous function  and of a subspace in $\bf Top$.
The `continuity' axiom~\eqref{eq:continuity} reflects the definition of a continuous function by means of the closure operator and
the `subspace regularity' axiom~\eqref{eq:subspace} reflects the definition of the induced topology on a subspace.

In all the examples below, the underlying factorization system is the epi/regular-mono factorization system.

\vskip 2mm \noindent{\bf Example 1.} As expected,
$\bf Top$ is a modal category: to see that subspaces are regular monos one may use for instance the two element set with the co-discrete topology as a regular subobject classifier.
%
As a modal category,
$\bf Top$ satisfies the following additional conditions:
\begin{itemize}
\item \emph{(Closure)} for every $X$ and $S \in \Sub_{\cM}(X)$
\begin{equation}\label{eq:closure}
 S \leq \Diamond S,\qquad \Diamond S = \Diamond \Diamond S;
\end{equation}
\item \emph{(Variable Independence)} for every $X_1, X_2$ and
      $S_1\in \Sub_{\cM}(X_1), S_2\in \Sub_{\cM}(X_2)$, we have
      \begin{equation}\label{eq:products}
       (\pi_{X_1}^* \Diamond S_1) \wedge
      (\pi_{X_2}^* \Diamond S_2) ~=~
      \Diamond (\pi_{X_1}^*  S_1 \wedge
      \pi_{X_2}^* S_2)~~.
      \end{equation}
\end{itemize}
These conditions are due to  specific properties of the closure operator and of the product topology
(variable independence implies in particular that the projection maps $\pi_{X_1},
\pi_{X_2}$ are \emph{open}
in the sense that $\pi_{X_1}^*,
\pi_{X_2}^*$ fully preserve the $\Diamond$ closure operator).
We shall turn to the question of whether the above conditions are sufficient for an
embedding theorem into $\bf Top$ in another paper (a partial positive result, for
function-free first-order one-sorted languages has been obtained in \cite{GhiModalitaCategorie1990,viareggio}).
$\hfill \dashv$

\vskip 2mm \noindent{\bf Example 2.} The category of Kripke frames $(W,R)$ and of stable (i.e. $R$-preserving) 
maps~\cite{rosalie}
is a modal category; so is the category $\bf POr$ of preordered sets and order preserving maps.
$\hfill \dashv$

\vskip 2mm \noindent{\bf Example 3.}
In
this paper, we concentrate on relational semantic environments similar to those
introduced in~\cite{GhiModalitaCategorie1990,viareggio,GMJSL} (see the Introduction Section~\ref{sec:intro} for additional references).
Given a graph
$$
{\bf G} :=  \{d,c: A \rightrightarrows O\}\text{,}
$$
a \emph{$\bf G$-relational set} $X$ consists of the following data:
\begin{itemize}
\item for every $\alpha\in O$, a set $X_{\alpha}$;
\item for every $k\in A$ with $d(k)=\alpha$ and $ c(k)=\beta$,  a relation
$X_k\subseteq X_{\alpha}\times X_{\beta}$.
\end{itemize}
Below, we write $k:\alpha\longrightarrow \beta$ to mean that $d(k)=\alpha, c(k)=\beta$ and $a\,k\,b$ to mean that $(a,b) \in X_k$.
A morphism of $G$-relational sets from $X$ to $Y$ is a family of functions
$$
f:= ~\{f_{\alpha}: X_{\alpha}\longrightarrow Y_{\alpha}\}_{\alpha \in O}
$$
that preserve the relations, i.e. such that
$$a\,k\,b \Rightarrow f_{\alpha}(a)\; k\; f_{\beta}(b)$$
for all $k:\alpha\longrightarrow \beta$, $a\in X_\alpha, b\in X_\beta$. We denote by $\cRel^{\bf G}$ the category of $G$-relational sets, with composition and the identities defined pointwise. This category has limits, also computed pointwise. The regular subobjects of a $\bf G$-relational set $X$ are the families
$$
f:= ~\{S_{\alpha}\subseteq X_{\alpha}\}_{\alpha \in O}
$$
with the relations $S_k$ obtained by taking the restrictions of the relations $X_k$
(they are classified by the two-element constant  $\bf G$-relational set with total relations).
Epimorphisms are pointwise surjective functions, so that ${\cRel^{\bf G}}$ is an \Fregular
(actually an \FBoolean) category.
 ${\cRel^{\bf G}}$ \emph { has a natural structure of modal category}: for $S\in \Sub_r(X)$ and $a\in X_{\alpha}$ we put
 \begin{equation}\label{eq:counterpart}
  a\in (\Diamond S)_{\alpha} ~~{\rm iff}~~{\rm there~are~}k, b~{~\rm s.t.~} akb ~{\rm and}~b\in S_{c(k)}~~.
 \end{equation}
The next proposition shows the peculiarity of this example.
$\hfill \dashv$
\vskip 2mm

\begin{proposition}\label{prop:quasitopos}
 ${\cRel^{\bf G}}$ is a quasi-topos.
\end{proposition}

\begin{proof}
 In fact ${\cRel^{\bf G}}$ can be presented as the category of separated
presheaves on a  site $({\bf C}, {\bf J})$ as follows (that separated presheaves are a quasi-topos is proved in \cite[Section A2.6]{elephant}). There are two kinds of objects in $\bf C$: for each vertex
$\alpha\in {\bf G}$, there is an object $[\alpha]$  in $\bf C$; for each edge $k:\alpha \longrightarrow \beta$ in  $\bf G$, there is an object $[k]$ in $\bf C$. For each edge $k:\alpha \longrightarrow \beta$ in  $\bf G$, there are two arrows $[\alpha]\longrightarrow [k]$ and $[\beta]\longrightarrow [k]$ in $\bf C$.  These are
the only non-identity arrows. The topology $\bf J$ consists of all the identity coverings, plus the
coverings of the form $[\alpha]\longrightarrow [k]$ and $[\beta]\longrightarrow [k]$ for each edge $k:\alpha \longrightarrow \beta$ in  $\bf G$.
These coverings satisfy the axioms for a Grothendieck topology (see \cite[Definition 2.1.9]{elephant}).

A presheaf $X : {\bf C}^{op} \longrightarrow {\bf  Set}$ is separated iff $X([k])\subseteq
X([\alpha])\times X([\beta])$ for each edge $k:\alpha \longrightarrow \beta$ in  $\bf G$.
 This means that $X$ is given by a family of sets $X([\alpha])$, plus a relation $X([k])$
for every edge $k$ in $\bf G$.
\end{proof}

\vskip 2mm \noindent{\bf Example 4.}
Given a category $\bf C$, we can consider relational presheaves~\cite{louvain,Niefield}: these are correspondences $X$ associating with every object $\alpha$ of $\bf C$ a set $X_{\alpha}$ and with every arrow $k:\alpha\longrightarrow \beta$
in $\bf C$ a relation $X_k\subseteq X_{\alpha}\times X_{\beta}$ in such a way that the lax functoriality conditions
$$
id_{\alpha} \subseteq X_{id_\alpha} \qquad X_k\circ X_l \subseteq X_{l\circ k}
$$
hold, for every $\alpha$ and every pair of composable arrows $k,l$. The arrows of the category $\cRel^{\bf C}$ of relational presheaves are defined in the same way as for ${\cRel^{\bf G}}$. It turns out that $\cRel^{\bf C}$ is a modal category, but it is not a quasi-topos anymore: quasi-toposes are regular categories~\cite{elephant} and for ${\bf C}$ equal to the singleton category $\bf 1$ we have $\cRel^{\bf 1}\simeq \cat{POr}$ which is not regular.
$\hfill \dashv$

\section{The Representation Theorem}\label{sec:representation}

Recall that a \emph{morphism} $\Phi$ between \Fcoherent categories $\Phi:({\cE_1}, \cM_1)\longrightarrow ({\cE_2}, \cM_2)$ is a
lex functor that preserves factorizations and joins; 
if $({\cE_1}, \cM_1)$ and $({\cE_2}, \cM_2)$ are modal categories,
a \emph{modal morphism} between them is an \Fcoherent functor $ \Phi: ({\cE_1}, \cM_1)\longrightarrow ({\cE_2}, \cM_2)$ that also preserves 
 the modal operators.
 
We say that $\Phi$ is a \emph{conservative} (resp. $\cM$-\emph{conservative}) \emph{embedding} when for any monomorphism $m$ (resp. $m\in \cM_1$), if $\Phi(m)$ is iso, then $m$ is iso as well.
For instance, the forgetful functor from $\bf Top$ to $\bf Set$ is an $\cM$-conservative embedding, but not a conservative embedding. We shall see that it is always possible to embed an \Fregular category $(\cE,\cM)$ into a power of $\bf Set$ in an $\cM$-conservative way, but not necessarily in a conservative way, because monos which are also in $\Mc^\bot$ become automatically isomorphisms in a topos like $\bf Set$ and its powers.

We shall prove a representation theorem of modal categories into categories of relational $\bf G$-sets: the construction will be very similar to the celebrated Joyal construction for the completeness theorem of intuitionistic first-order logic~\cite[p.~75]{MR}.
The representation will take the form of an $\cM$-conservative embedding, to be strengthened to a conservative embedding, in the case of modal categories satisfying the saturation condition
(see Section~\ref{sec:comprehension} below).
In order to prove such a theorem, we first fix some notation about models in $\bf Set$ and review some basic material concerning Lawvere doctrines.

A \emph{coherent model} (or just a \emph{model})
of an f-coherent category $(\cE, \cM)$ is an f-coherent functor to $\bf Set$ equipped with its usual factorization system of surjections and injections. The 
class of the
%
models of $(\cE, \cM)$
is denoted by $\cMod(\cE,\cM)$. Thanks to Proposition~\ref{prop:reg-full-hyperdoc} which transfers to f-coherent categories and coherent Lawvere doctrines, these models coincide with the models of the associated coherent Lawvere doctrine. 

To make notation easier, for a given model $M : {\cE} \superto {\bf Set}$, we shall often use $X_M, f_M, {\ldots}$ instead of $M(X), M(f),{\ldots}$ to indicate the values of the functor $M$ on objects and arrows of $\cE$.
For a subobject $S\buildrel s\over\hookrightarrow X$ in $\Sub_{\cM}(X)$, we might use also the notation 
$\lbrack\!\lbrack S\rbrack\!\rbrack_M$ for the subset of $M(X)$ given by image of the function $M(S) \buildrel{M(s)}\over\hookrightarrow M(X)$ (this is the subset of $M(X)$ canonically representing the subobject 
$M(S) \buildrel{M(s)}\over\hookrightarrow M(X)$). Equalities like 
 $\lbrack\!\lbrack\exists_f S\rbrack\!\rbrack_M= \exists_{f_M}\lbrack\!\lbrack S\rbrack\!\rbrack_M$,
 and $\lbrack\!\lbrack f^* T\rbrack\!\rbrack_M= f_M^*\lbrack\!\lbrack T\rbrack\!\rbrack_M$ are immediate consequences of our definitions and notational conventions (they will be frequently used in this section).

We need a model existence theorem for Lawvere doctrines, to be applied to doctrines of the kind $({\cE}, \Sub_{\cM})$ for an \Fcoherent category $({\cE}, \cM)$. Recall that a \emph{filter} in a distributive lattice $(D, \wedge, \top, \vee, \bot)$ is a subset $F\subseteq D$ that contains $\top$, is closed under $\wedge$ and is upward-closed relatively to the ordering $\leq$ of the lattice. Dually, an \emph{ideal} is a downward-closed subset containing $\bot$ and closed under $\vee$. A filter $F$ is \emph{prime} if it does not contain $\bot$ and contains either $x$ or $y$ in case it contains $x\vee y$. The existence of enough prime filters is a consequence of the following well-known lemma (whose validity depends on choice axiom):

\begin{lemma}\label{lem:ee}
If $\Gamma$ and $\Delta$ be two disjoint subsets of a distributive lattice $D$ such that $\Gamma$ is a filter and $\Delta$ is an ideal, then there exists a prime filter $P$ such that $P\supseteq \Gamma$ and $P\cap \Delta=\emptyset$.
\end{lemma}

Let $({\cE}, \Pc)$ be a coherent Lawvere doctrine and let $X$ be an object of $\cE$; a prime filter $P$ of $\Pc(X)$ is called an \emph{$X$-type} of $\Pc$. Typical $X$-types come from models: if $(M,\alpha)$ is a model and $x\in M(X)$, the set $P$ of the $S\in \Pc(X)$ such that $x\in \alpha_X(S)$ is a type; we say in this case that \emph{$x$ realizes $P$}. All types can in fact be realized, as stated in the following lemma, which is nothing but a small variant of the G\"odel completeness theorem for first order logic (see e.g.~\cite[Thm~7.2]{vanMarDualityModelTheory2024}):

\begin{proposition}\label{prop:real}
  Given a a coherent Lawvere doctrine $({\cE}, \Pc)$ and an $X$-type $P$ of $\Pc$, there is a model $(M,\alpha)$ of $({\cE}, \Pc)$ and an element $x\in M(X)$ realizing $P$.
\end{proposition}

Proposition~\ref{prop:real} can be used to prove a representation theorem for small \Fcoherent categories
into powers of $\bf Set$ as follows. Notice that given any \Fcoherent functor $\Phi: ({\cE}, \cM) \lra {\bf Set}^W$ from $({\cE}, \cM)$ into a $W$-power of $\bf Set$ and given $w\in W$, the composite functor
$$
ev_w\circ \Phi: ({\cE}, \cM) \lra {\bf Set}^W \lra {\bf Set}
$$
 is a \Fcoherent model of $({\cE}, \cM)$ (here $ev_w$ is the ``evaluation at $w$'' functor, i.e. the functor induced by the inclusion $\{w\}\hookrightarrow W$). If we want to show that
$({\cE}, \cM)$ has an $\cM$-conservative embedding into a power of $\bf Set$, the obvious idea is to take $W$ to be the class $\cMod({\cE}, \cM)$ of all the \Fcoherent models of $({\cE}, \cM)$.\footnote{One may complain that the set $\cMod({\cE}, \cM)$ is not small: however, this problem can be solved in one of the ways which are familiar to logicians (for instance it is sufficient to limits to models taking values into sets which are in fact subsets of a preassigned set whose cardinality is bigger or equal than the cardinality of $\cE$).}
 We are lead to the following

\begin{theorem}\label{thm:rep} For every small \Fcoherent category $(\cE, \cM)$,
 the evaluation functor
 $$
 ev_{({\cE}, \cM)}= (ev_M \mid M\in \cMod({\cE}, \cM)) : {\cE} \longrightarrow {\bf Set}^{\cMod({\cE}, \cM)}
$$
 is a $\cM$-conservative embedding of \Fcoherent categories.
\end{theorem}

\begin{proof}
 Trivially, limits and factorizations are all preserved because they are computed pointwise in powers of $\bf Set$. $\cM$-conservativity (but not conservativity!)
 is ensured by Proposition~\ref{prop:real}: if $S\buildrel{s}\over\hookrightarrow X$ is in $\cM$ but is not and iso, then $S\neq \top$ holds in $\Sub_{\cM }(X)$, hence there is an $X$-type $P$ such that $S\not \in P$ by Lemma~\ref{lem:ee}. If $P$ is realized in $M$ by some $a\in M(X)$, we have that
 $a\not \in \lbrack\!\lbrack S\rbrack\!\rbrack_M$, which means that the image of $M(s)$ is not $M(X)$.
 Thus $ev_{({\cE}, \cM)}(s)$ is not iso because its $M$-component is not such.
\end{proof}

In order to extend the above machinery to modal categories, we need a notion of a modal transformation between two
\Fcoherent models of a modal category. This notion will be the modal analogue of a natural transformation --- aka elementary embedding --- among two models of a classical first order theory.
We again take inspiration from modal morphisms of the kind $\Phi:({\cE}, \cM)\lra {\cRel^{\bf G}}$. Notice that
for every $k:\alpha\longrightarrow \beta$ in $\bf G$ such a modal morphism $\Phi$ produces a composite functor
$$
({\cE}, \cM)~\buildrel{\Phi}\over\lra~ {\cRel^{\bf G}}~\lra~ {\bf Rel}^{(\alpha \buildrel{k}\over\rightarrow \beta)}
$$
 by composition with the functor induced by the sub-graph
inclusion $(\alpha \buildrel{k}\over\rightarrow\nobreak \beta)\allowbreak \hookrightarrow \bf G$. This composite functor
represents $k$ as a family of relations (indexed by the objects of $\cE$) between two coherent models of $({\cE}, \cM)$.
Such a family of relations satisfies the following definition, thus motivating it.

Formally, we define a \emph{modal transformation} $R: M\longrightarrow N$ between two coherent models
$$
M: ({\cE}, \cM) \longrightarrow {\bf Set}~~{\rm and}~~
N: ({\cE}, \cM) \longrightarrow { \bf Set}
$$
of a modal category $(\cE, \cM)$ as a lex subfunctor
$$
R\subseteq M\times N
$$
satisfying the `reflection' condition
\begin{equation}\label{eq:refl}
s_M(a) \; R(X) \; s_N(b) ~~\Rightarrow~~ a \; R(S) \; b
\end{equation}
(for all $S\buildrel{s}\over \hra X\in \cM$, $a\in S_M, b\in S_N$)
and the `continuity' condition:
\begin{equation}\label{eq:cc}
\Diamond_{R(X)} \lbrack\!\lbrack S\rbrack\!\rbrack_N \;\subseteq\; \lbrack\!\lbrack \Diamond S\rbrack\!\rbrack_M~~.
\end{equation}
(for all $X$ and $S\in \Sub_{\cM}(X)$).

\begin{remark}
 Unraveling condition~\eqref{eq:cc} according to
~\eqref{eq:diamond},\eqref{eq:diamond1}, we can reformulate it as
\begin{equation}\label{eq:ccexpl}
a \,R(X)\,b ~{\rm and}~b\in \lbrack\!\lbrack S\rbrack\!\rbrack_N
~~\Rightarrow~~ a\in \lbrack\!\lbrack \Diamond S\rbrack\!\rbrack_M
\end{equation}
(for all $a\in X_M$ and $b\in X_N$). This is clearly reminiscent of the definition of the accessibility relation in canonical models widely used in the modal logic literature~\cite{CZ}.
\end{remark}

\begin{remark}
 We defined a modal transformation to be a lex subfunctor (i.e. a subfunctor preseving finite products
 and equalizers) satisfying reflection and continuity conditions. This definition is slightly redundant, because reflection implies preservation of equalizers (recall that equalizers are in $\cM$). On the other hand, if $\cM$ is the class of regular monos (as it often happens in the semantic examples), then reflection becomes 
 redundant.
\end{remark}

We let $\bf{G(E,\cM)}$ be the (large) graph of all the coherent models of a modal category $({\cE}, \cM)$ and of all the modal transformations among them.
We have an evaluation functor
$$
 ev_{({\cE}, \cM)}: {\cE} \longrightarrow {\bf Rel^{G(E, \cM)}}
$$
mapping an object $X$ in $\cE$ to the $\bf{G(E,\cM)}$-relational set $ev_{({\cE}, \cM)}(X)$ whose $M$-component is $X_M$ and whose $R$-component (for $R: M\longrightarrow N$ in $\bf G(E,\cM)$) is $R(X)$.
The action of the functor $ev_{({\cE}, \cM)}$ on an arrow $f:X\longrightarrow Y$ is defined by $ev_{({\cE}, \cM)}(f)(M) = f_M$. Our representation theorem now reads as:

\begin{theorem}\label{thm:main} For every small modal category $({\cE}, \cM)$,
 the evaluation functor
 $$
 ev_{({\cE}, \cM)}: ({\cE}, \cM) \longrightarrow {\bf Rel^{G(E,\cM)}}
$$
 is a $\cM$-conservative embedding of modal categories.
\end{theorem}

The proof of Theorem~\ref{thm:main} occupies the remaining part of this section.
We fix a small modal category $(\cE, \cM)$.
We spell out in detail the conditions for $R$ to be a modal transformation between two models
$M$ and $N$ of $\cE$. Below we often omit types for simplicity and write simply $a R b$ when $a\in X_M, b\in Y_N$ and $(a,b) \in R(X)$; similarly, we write just $f(a)$
for $f_M(a)$, when $f:X\longrightarrow Y$ is an arrow in $\cE$ and $a\in X_M$.
First of all, $R$ is in fact a family of relations $R=\{ R(X)\subseteq X_M\times X_N\}_X$ indexed by the objects of $\cE$;
the fact that such a family $R$ is a lex subfunctor and satisfies reflection means that the following conditions are satisfied:
\begin{eqnarray}
 &
 a_1Rb_1 \;\& \cdots \&\; a_nRb_n \Rightarrow (a_1, \dots, a_n) R (b_1, \dots, b_n) ~({\rm for}~n\geq 0)\label{eq2}
 \\ &
 a Rb \Rightarrow f(a) R f(b) ~~({\rm for~all}~f~{\hbox{\rm of appropriate domain in}}~\cE)
 \label{eq1}
 \\ &
 s(a) R s(b) \Rightarrow a Rb,~~({\rm for}~ S\buildrel{s}\over \hookrightarrow X \in \Sub_{\cM}(X)) \label{eq3}
\end{eqnarray}
In addition we have condition~\eqref{eq:cc}, namely
\begin{equation}\label{eq4}
 aRb ~{\rm and}~b\in \lbrack\!\lbrack S\rbrack\!\rbrack_N
~~\Rightarrow~~ a\in \lbrack\!\lbrack \Diamond S\rbrack\!\rbrack_M~~.
\end{equation}

We will sometimes write $X_R$ instead of $R(X)$, using the same convention for modal transformations as for models.

An \emph{$\cM$-partial map}
is a span $\begin{tikzcd}
		X &[-0.5em] S \ar[l,hook'] \ar[r," "] &[-0.5em] Y
	\end{tikzcd}$ where $S\hookrightarrow X$ is in $\cM$.

\begin{lemma}\label{lem:gen1}
Given two models $M,N$ and a family of relations $R=\{ R(X)\subseteq M(X)\times N(X)\}_X$ indexed by the objects of $\cE$, the least family of relations $\tilde R$ containing $R$ and satisfying conditions~\eqref{eq1}, \eqref{eq3} can be described as follows.

For $a\in Y_M, b\in Y_N$ we have $a {\tilde R}b$ iff there exists a span
\[\begin{tikzcd}
  Y & S \ar[l,"f"'] \ar[r,hook,"s"] & X
\end{tikzcd}\]
where $s \in \Mc$ and $a' \in S_M, b' \in S_N$ such that:
\begin{itemize}
  \item $f(a') = a$ and $f(b') = b$;
  \item $s(a')~R~s(b')$.
\end{itemize}
\end{lemma}

\begin{proof}
The proof is a simple check, using the stability of $\cM$ under pullbacks and the fact arrows in $\cM$ compose.
\end{proof}


\begin{lemma}\label{lem:gen2}
 If, for two given models $M,N$, a family $R$ of relations indexed by the objects of $\cE$ satisfies condition~\eqref{eq4}, then so does the family of relations $\tilde R$ mentioned in Lemma~\ref{lem:gen1}.
\end{lemma}

\begin{proof}
 This is due to the continuity and subspace conditions of Definition~\ref{def:main}. Suppose in fact that we have $a {\tilde R}b$ so that there exists the data mentioned in the statement of Lemma~\ref{lem:gen1}. Suppose that $b\in \lbrack\!\lbrack U\rbrack\!\rbrack_N$ for some $U\buildrel{u}\over{\hookrightarrow} Y$. From
 $f(b')=b$ we obtain
 $$
 b' \in f_N^*\lbrack\!\lbrack U\rbrack\!\rbrack_N
 = \lbrack\!\lbrack f^* U\rbrack\!\rbrack_{N}~~;
 $$
 thus
 $$
 s(b') \in \exists_{s_N} \lbrack\!\lbrack f^* U\rbrack\!\rbrack_{N}=
 \lbrack\!\lbrack \exists_{s} f^* U\rbrack\!\rbrack_{N}~~.
 $$
 Since $R$ satisfies~\eqref{eq4} and $s(a')~R~s(b')$, we obtain
 $$
 s(a') \in
 \lbrack\!\lbrack \Diamond\exists_{s} f^* U\rbrack\!\rbrack_{M}~~.
 $$
 and
 $$
 a' \in {s}_M^* \lbrack\!\lbrack \Diamond\exists_{s} f^* U \rbrack\!\rbrack_{M}=
 \lbrack\!\lbrack s^*\Diamond\exists_{s} f^* U \rbrack\!\rbrack_{M}~~.
 $$
 By subspace and continuity conditions from Definition~\ref{def:main}, we have
 $$
 a' \in \lbrack\!\lbrack s^* \Diamond\exists_{s} f^* U\rbrack\!\rbrack_{M}=
 \lbrack\!\lbrack \Diamond f^* U\rbrack\!\rbrack_{M} \subseteq
 \lbrack\!\lbrack f^*\Diamond U\rbrack\!\rbrack_{M}=
 f_M^* \lbrack\!\lbrack \Diamond U\rbrack\!\rbrack_{M}~~.
 $$
 Now, $f(a')=a$ yields $a\in \lbrack\!\lbrack \Diamond U\rbrack\!\rbrack_{M}$, as desired.
\end{proof}

In the following lemma, we say that the family of relations $R$ is a \emph{singleton} if only one of the $R(X)$ is non-empty and that this one contains a single element.

\begin{lemma}\label{lem:gen3}
    Let $M,N$ be two models of $\cE$ and let $R$ be a family of relations indexed by the objects of $\cE$. Suppose that either $R$ satisfies \eqref{eq2}, or that $R$ is a singleton. Then the family $\tilde{R}$ defined in Lemma~\ref{lem:gen1} satisfies \eqref{eq2}.
\end{lemma}

\begin{proof}
    Suppose that $R$ satisfies \eqref{eq2}. The only element of 
    ${\bf 1}_M \times {\bf 1}_N$ (here $\bf 1$ is the terminal object of $\cE$) is in $\tilde{R}$ because it is in $R$. The closure of $\tilde{R}$ under binary products is seen by taking the product of two diagrams of the shape presented in Lemma~\ref{lem:gen1} and by using that $s_1 \times s_2 \in \Mc$ whenever $s_1, s_2 \in \Mc$.

    Suppose that $R$ is a singleton and contains only $(a,b) \in X_M \times X_N$. Let $R_p$ be the closure of $R$ under \eqref{eq2}. We first show that $R_p \subseteq \tilde{R}$. Indeed, for any $n\geq 0$, the image of $(a,b)$ by the 
    diagonal
    map $X_M \times X_N \to X_M^n \times X_N^n$ is in $\tilde{R}$, and $R_p$ consists precisely of all these elements. Since $R_p \subseteq \tilde{R}$, we obtain that $\tilde{R_p} = \tilde{R}$ because $\tilde{R_p} \subseteq \tilde{\tilde{R}} = \tilde{R}$. Hence $\tilde{R_p}$ satisfies \eqref{eq2} by the first part of this lemma.
\end{proof}

We can now conclude the {\bf proof of Theorem~\ref{thm:main}}:

\begin{proof}
We must prove that the evaluation functor is $\cM$-conservative and preserves all the operations. $\cM$-conservativity is proved in the same way as in Theorem~\ref{thm:rep}.

{\bf $ev_{({\cE}, \cM)}$ preserves finite limits.} For any coherent model $M$, the functor $X \mapsto M(X)$ preserves finite limits. For any modal transformation $R \subseteq M\times N$, the functor $X \mapsto R(X)$ preserves finite limits. Since limits in
${\bf Rel^{G(E,\cM)}}$ are computed pointwise,
this shows that $ev_{({\cE}, \cM)}$ preserves finite limits.

{\bf $ev_{({\cE}, \cM)}$ preserves the left class of the factorization system.} The left class of the factorization system of ${\bf Rel^{G(E,\cM)}}$ consists of all the morphisms whose $M$-component is surjective for each coherent model $M$. If $f : X\lra Y$ is a morphism in the left class of $\cE$, then $M(f) : M(X) \lra M(Y)$ is surjective for each coherent model $M$. We get that $ev_{({\cE}, \cM)}(f)$ is in the left class of ${\bf Rel^{G(E,\cM)}}$.

{\bf $ev_{({\cE}, \cM)}$ preserves the right class of the factorization system.} The right class of the factorization system of ${\bf Rel^{G(E,\cM)}}$ consists of all the morphisms $g : V\lra W$ such that $g_M : V_M \lra W_M$ is injective for each coherent model $M$, and moreover such that for each modal transformation $R \subseteq M\times N$ between coherent models, $V_R \subseteq V_M \times V_N$ is the restriction of $W_R \subseteq W_M \times W_N$ via $g_M \times g_N$. If $f:X\lra Y$ is a morphism in the right class $\cM$ of $\cE$, then $f_M:X_M \lra Y_M$ is injective for each coherent model $M$. Moreover, the reflection condition~\eqref{eq3} for $R$ imposes that $X_R \subseteq X_M\times X_N$ is the restriction of $Y_R \subseteq Y_M\times Y_N$ via $f_M\times f_N$. We obtain that $ev_{({\cE}, \cM)}(f)$ is in the right class of ${\bf Rel^{G(E,\cM)}}$.

{\bf $ev_{({\cE}, \cM)}$ preserves the modality.} One side of the preservation condition is~\eqref{eq4}, the other side is a converse of~\eqref{eq4}, namely
\begin{equation}\label{eq5}
a\in \lbrack\!\lbrack \Diamond S\rbrack\!\rbrack_M
~~\Rightarrow~~{\hbox{\rm there are}}~N, b, R~{\rm s.t.}~
 aRb ~{\rm and}~b\in \lbrack\!\lbrack S\rbrack\!\rbrack_N
\end{equation}
for every $M$, $X$, $S\in \Sub_{\cM}(X)$ and $a \in X_M$. To prove it,
pick such $a\in \lbrack\!\lbrack \Diamond S\rbrack\!\rbrack_M$ and consider the following subsets of $\Sub_{\cM}(X)$:
\begin{equation}\label{eq6}
\Gamma~=~\{ S\}, \qquad \Delta~=~\{ T\mid a\not\in \lbrack\!\lbrack \Diamond T\rbrack\!\rbrack_M\}
\end{equation}
Clearly $\Gamma$ is closed under finite meets and $\Delta$ is closed under finite joins
(because $\Diamond$ commutes with joins) and downward-closed (because $\Diamond$ preserves the lattice ordering); in addition $\Gamma\cap \Delta=\emptyset$ since $a\in \lbrack\!\lbrack \Diamond S\rbrack\!\rbrack_M$. By Proposition~\eqref{prop:real}, there are a model $N$ and $b\in X_N$
realizing a prime filter extending $\Gamma$ and disjoint from $\Delta$. This means in particular that $b\in \lbrack\!\lbrack S\rbrack\!\rbrack_N$ and that the family of relations $R=\{ R(Y)\subseteq Y_M\times Y_N\}_Y$ containing just the singleton pair
$(a, b) \in R(X)$ satisfy condition~\eqref{eq4}. Now it is sufficient to apply Lemmas~\ref{lem:gen2} and \ref{lem:gen3}.
\end{proof}

\section{Saturation}\label{sec:comprehension}

Saturation aspects must be carefully addressed in a modal category 
$({\bf E}, \cM)$.
Here is a typical saturation problem:

\begin{quote}
    Suppose that $R\hookrightarrow X\times Y$ in $\Mc$ is functional (i.e.\ that it satisfies the axioms of a functional relation in the Lawvere doctrine $\Sub_\Mc$ of \S~\ref{subsec:hyperdoc}) and that it is `continuous' in the obvious sense that `inverse image along $R$' satisfies condition~\eqref{eq:continuity}. Is $R$ necessarily the graph of an arrow $X\lra Y$?
\end{quote}

In general, there is no reason for the answer to be positive. Analogous problems arise concerning the notion of a subspace and of an isomorphism. We shall formulate below all the related saturation principles and then we shall prove that they are all equivalent, giving rise to a stronger notion of a `saturated' modal category. 

Let us start by analyzing the notion of a subspace, i.e.\ of an arrow 
$m:A \hookrightarrow X$ belonging to $\cM$. We know that $m$ must satisfy condition~\eqref{eq:subspace}. We wonder what happens if a converse is satisfied: namely, is it sufficient that~\eqref{eq:subspace} holds for all $S\in \Sub_{\cM}(A)$ for a \emph{monomorphism} $m$ to be in $\cM$? If we look at our reference semantic category ${\cRel^{\bf G}}$, we see that this is not the case, in the sense that we must make the condition `stable' by adding a parameter $Z$ for this to be true. This is similar to what happens for instance in Lawvere's doctrines \cite{hyperdoctrines} for the substitutivity axiom for equality: a parameter must be added to formulate it correctly. We say that a mono $m:A \hookrightarrow X$ is a \emph{brittle subspace} iff for all $Z$, for all $S\in \Sub_\Mc(A\times Z)$ we have that
\begin{equation}\label{eq:embedding}
\Diamond S=(m\times 1_Z)^* \Diamond\, \exists_{m\times 1_Z} \,S.
\end{equation}
In $\cRel^{\bf G}$, the $\Mc$-subobjects can then be characterized as the brittle subspaces.
\footnote{
In order to show that, $Z$ can be usefully instantiated in ${\cRel^{\bf G}}$ to relational $\bf G$-sets which are  `singleton pairs', i.e. on relational $\bf G$-sets $X$ that for some $k:\alpha\to \beta$ in $\bf G$, we have that $X_{\alpha}=\{*\}=X_\beta$ and that $X_h$ is empty unless $h=k$, in which case $X_K=\{(*,*)\}$.}
We can now formulate the

\vskip 2mm\noindent
{\bf Subspace saturation principle (SS).} \emph{A modal category satisfies
the subspace saturation principle iff every brittle subspace 
is a subspace (i.e.\ it belongs to $\cM$).}
\vskip 2mm

Now, let us analyze \emph{functional relations}: we define them to be the $R\buildrel{r}\over \hookrightarrow X\times Y$ in $\cM$ such that the composite
map
$$
R \buildrel{r}\over{\hookrightarrow} X\times Y \buildrel{\pi_X} \over \longrightarrow X
$$
is a monomorphism and is also in $\Mc^\bot$ (we shall see below how to reformulate this and other similar conditions at the doctrinal level, so as to be able to use the internal language of doctrines in computations). 

A \emph{continuous relation} on the other hand is some $R\buildrel{r}\over \hookrightarrow X\times Y$ in $\cM$ such that
for all $Z$, for all $S\in \Sub_{\cM}(Y\times Z)$ we have that
the following inequality holds in $Sub_{\cM}(X\times Y\times Z)$:
\begin{equation}\label{eq:Rcont0}
\Diamond \exists_{\pi_{X\times Z}}( \pi_{X\times Y}^*(R)\wedge \pi_{Y\times Z}^*(S)) \leq  \exists_{\pi_{X\times Z}}( \pi_{X\times Y}^*(R)\wedge \pi_{Y\times Z}^*(\Diamond S))~~.
\end{equation}
If, in general, we abbreviate
$\exists_{\pi_{X\times Z}}( \pi_{X\times Y}^*(R)\wedge \pi_{Y\times Z}^*(T))$
(for $T\in \Sub_{\cM}(Y\times Z)$)
as $R\circ T$, the above continuity condition can be written simply  as
\begin{equation}\label{eq:Rcont}
\Diamond (R\circ S) \leq R\circ (\Diamond S)~~.
\end{equation}
We say that $R\in Sub_\cM(X\times Y)$ is a \emph{brittle morphism} iff it is functional and
continuous.
We can now formulate the

\vskip 2mm\noindent
{\bf Functional saturation principle (FS).} \emph{
A modal category satisfies 
the functional saturation principle iff every brittle morphism $R\hookrightarrow X\times Y$ is the graph of an arrow, meaning that there is $h: X\lra Y$ such that we have $(1_X, h)\simeq R$
in $\Sub_{\cM}(X\times Y)$.
}

Finally, let us analyze the case of isomorphisms.
We say that $h: X\lra Y$ is \emph{open} iff it satisfies the condition
$$
h^*(\Diamond S) \leq \Diamond (h^* S)
$$
for every $S\in \Sub_{\cM}(Y)$; it is \emph{stably open} iff $h\times 1_Z$ is open for every $Z$. We say that $h$ is a \emph{brittle isomorphism} iff it is mono, it is in $\cM^\bot$ and it is also stably open. 

We can now formulate the

\vskip 2mm\noindent
{\bf Isomorphism saturation principle (IS).} \emph{A modal category satisfies the isomorphism saturation principle iff every brittle isomorphism is an isomorphism}.
\vskip 2mm

Before attacking the proof that the above saturation principles are all equivalent, let us make a practical remark. In concrete computations, it is convenient to rely on the doctrinal formalism and work in the underlying Lawvere doctrine of a modal category (see Proposition~\ref{p31.1}). To this aim it is important to know that most notions concerning an \Fcoherent category $({\bf E}, \cM)$ have equivalent reformulations at the doctrinal level in terms of subobjects inclusions. In particular, the fact that an arrow $f:X\lra Y$ belongs to $\Mc^\bot$ can be rewritten as $\top \leq \exists_f(\top)$ and the fact that $f$ is mono can be rewritten as $(f\times f)^*(=_Y) \leq (=_X)$, where $=_X, =_Y$ are the equalities over $X$ and $Y$ --- recall that the diagonals are in $\cM$, being the equalizers of the two projections.
Thus for instance, we can establish that $f^*$ is injective (with $\exists_f$ as left inverse) when $f\in \Mc^\bot$, that $f^*$ is surjective (with $\exists_f$ as right inverse) when $f$ is mono, that $\exists_f$ and $f^*$ are inverse to each other when $f\in \Mc^\bot$ is also mono, etc. We shall refer to these kinds of arguments as `first-order reasoning' below (such reasoning can be profitably worked off using the internal language of Lawvere doctrines).

\begin{theorem}\label{thm:saturation} The above saturation principle are all equivalent in a  modal category
$({\bf E}, \cM)$.
\end{theorem}

\begin{proof}

 (SS $\Rightarrow$ IS): Suppose that $h:X\lra Y$ is in $\Mc^\bot$, is mono and stably open. Take $Z$ and $S\in \Sub_\cM (X\times Z)$; we have
 $$
 \Diamond S ~=~\Diamond (h\times 1_Z)^* \exists_{h\times 1_Z} S~=~(h\times 1_Z)^*\Diamond \exists_{h\times 1_Z} S
 $$
 by first-order reasoning and stable openness. Thus $h\in \cM$ by (SS) and consequently $h$ is iso.

 (IS $\Rightarrow$ SS): Let $m:A\lra X$ be a brittle subspace; let us factor it as 
 $A \buildrel q\over \lra Q \buildrel i \over \lra X$
 with $i\in \cM$ and $q\in \Mc^\bot$.
 We show that $q$ is stably open (so that (IS) applies, $q$ is iso and then $m\in \cM$). Take $Z$ and 
  $S\in \Sub_\Mc(Q\times Z)$; 
 we have by~\eqref{eq:subspace}
 \begin{align*}
     (q\times 1_Z)^*\Diamond S &= (q\times 1_Z)^*(i\times 1_Z)^*\Diamond \exists_{i\times 1_Z} S \\
     &= (m\times 1_Z)^*\Diamond \exists_{i\times 1_Z} S \\
     &= (m\times 1_Z)^*\Diamond \exists_{i\times 1_Z} \exists_{q\times 1_Z} (q\times 1_Z)^* S \tag{first-order reasoning}\\
     &= (m\times 1_Z)^*\Diamond \exists_{m\times 1_Z} (q\times 1_Z)^* S \\
     &= \Diamond (q\times 1_Z)^* S \tag{by the hypothesis on $m$}
 \end{align*}
 This shows that $q\times 1_Z$ is open.

 (FS $\Rightarrow$ IS): Let $h:X\lra Y$ be a brittle isomorphism; consider $R$ defined as
 $$
 R:= X\buildrel{(h, 1_X)}\over\lra Y\times X
 $$
 Notice that $\pi_Y\circ R=h$ is mono and belongs to $\Mc^\bot$ by our hypotheses on $h$, so $R$ is a functional relation; if we manage to prove that $R$ is also 
 continuous, then by functional saturation there is $f:Y\lra X$ such that $R=(h,1_X) \simeq (1_Y,f)$, which means that $h$ is iso with inverse $f$.

 Now notice that, by first-order reasoning, for every  $Z$ and $S\in \Sub_{\cM}(X\times Z)$ we have that
 $R\circ S$ is equal to $\exists_{h\times 1_Z}(S)$. As a consequence, the continuity of our $R$ can be expressed as the ``Barcan formula'' for $h\times 1_Z$
 $$
 \Diamond \exists_{h\times 1_Z}S \leq \exists_{h\times 1_Z} \Diamond S,
 $$
 which is an immediate consequence of the stable openness of $h$ and the fact that $h$ (and so also $h\times 1_Z$) satisfies the hypotheses of  Lemma~\ref{lem:BF} below.

 (IS $\Rightarrow$ FS): Let $R\buildrel{r}\over \hookrightarrow X\times Y$ be functional and continuous. We show that $R\buildrel{r}\over \hookrightarrow X\times Y \buildrel{\pi_X}\over \lra X$ is stably open: then (IS) applies and $\pi_X\circ r$ has an inverse $h:X\longrightarrow R$ witnessing the isomorphism of $R$ with the graph of $\pi_Y\circ r\circ h$.
 In order to show that $\pi_X\circ r$ is stably open, we use Lemma~\ref{lem:BF} and prove the ``Barcan formula'' for
 $(\pi_Y\circ r) \times 1_Z$ for every $Z$. To this aim, let $S\in \Sub_{\cM}(R\times Z)$; we have
 $$
 \Diamond \exists_{(\pi_X\circ r)\times 1_Z} S~=~
 \Diamond \exists_{\pi_{X\times Z}}\exists_{r\times 1_Z} S~ =~ \Diamond \exists_{\pi_{X\times Z}} (\pi_{X\times Y}^*R\wedge \exists_{r\times 1_Z} S)
 $$
 by first-order reasoning and
 $$
  \Diamond \exists_{\pi_{X\times Z}} (\pi_{X\times Y}^*R\wedge \exists_{r\times 1_Z} S)
 ~\leq~
 \exists_{\pi_{X\times Z}} (\pi_{X\times Y}^*R\wedge \Diamond \exists_{r\times 1_Z} S)
 $$
 by Lemma~\ref{lem:contZ} below. We now continue by first-order reasoning\footnote{Notice that $(\pi_{X\times Y}^*R)\wedge U=\exists_{r\times 1_Z} (r\times 1_Z)^*(U)$ for every 
 $U\in Sub_{\cM}(X\times Y\times Z)$. 
 }
   and obtain
 $$
 \exists_{\pi_{X\times Z}} (\pi_{X\times Y}^*R\wedge \Diamond \exists_{r\times 1_Z} S)
 = \exists_{\pi_{X\times Z}} \exists_{r\times 1_Z}({r\times 1_Z})^*\Diamond \exists_{r\times 1_Z} S
 $$
 which is equal to $\exists_{(\pi_X\circ r)\times 1_Z}\Diamond S$,
 as required.
\end{proof}

\begin{lemma}\label{lem:BF}
Suppose that $h\in \Mc^\bot$ is also mono. Then $h$ is open iff it satisfies the ``Barcan formula'', namely
 \begin{equation}\label{eq:BF}
  \Diamond \exists_{h} S \leq \exists_{h} \Diamond S~~.
 \end{equation}
 for every $\cM$-subobject $S$ of the domain of $h$.
\end{lemma}

\begin{proof}
 Trivial, by first-order reasoning, taking in mind that $\exists_h$ is the inverse correspondence of $h^*$.
\end{proof}

\begin{lemma}\label{lem:contZ}
Suppose that $R\in \Sub_{\cM}(X\times Y)$ is continuous; take $\tilde Z$ and $U\in
\Sub_{\cM}(Y\times X\times {\tilde Z})$. We have
\begin{equation}\label{eq:contZ-target}
    \Diamond \exists_{\pi_{X\times \tilde Z}} (\pi_{X\times Y}^*R\wedge U)
 ~\leq~
 \exists_{\pi_{X\times \tilde Z}} (\pi_{X\times Y}^*R\wedge \Diamond U)~~.
\end{equation}
\end{lemma}

\begin{proof}
 Observe that, by 
 first-order reasoning or directly by applying the Beck-Chevalley condition to a suitable pullback square, we have
 \begin{equation}\label{eq:bc}
  \exists_{\pi_{X\times \tilde Z}}((\pi_{X\times Y}^*R)\wedge U) = (\Delta_X \times 1_{\tilde Z})^*(R\circ U) \text{.}
 \end{equation}
 Thus we get
 \begin{align*}
     \Diamond \exists_{\pi_{X\times \tilde Z}} (\pi_{X\times Y}^*R\wedge U) &= \Diamond (\Delta_X \times 1_{\tilde Z})^*(R\circ U) \\
     &\leq (\Delta_X \times 1_{\tilde Z})^* \Diamond (R\circ U) \\
     &\leq (\Delta_X \times 1_{\tilde Z})^* (R \circ \Diamond U) \tag{by continuity of $R$} \\
     &= \exists_{\pi_{X\times \tilde Z}} (\pi_{X\times Y}^*R\wedge \Diamond U)
 \end{align*}
\end{proof}

\begin{definition}
 We say that a modal category is \emph{saturated} iff it satisfies one of (hence all) the saturation principles {\rm (SS),(FS),(IS)}.
\end{definition}

An effect of the saturation condition is that  $\cM$-conservative embeddings coincide with conservative embeddings.

\begin{proposition}\label{prop:conservativity}
 Let $\Phi:({\bf E_1}, \cM_1)\longrightarrow ({\bf E_2}, \cM_2)$ be a modal morphism and let $({\bf E_1}, \cM_1)$ be a  saturated modal category; then $\Phi$ is $\cM$-conservative
 iff it is conservative.
\end{proposition}

\begin{proof}
 The right to left direction is trivial.
 For the converse, let  $\Phi$ be $\cM$-conservative. Notice that, for $S,T\in \Sub_{\cM_1}(X)$,  we have that
 ``$S\leq T$ and $\Phi(S)\simeq \Phi(T)$ imply $S\simeq T$'': this is because of Lemma~\ref{lem:firstcomponent} and $\cM$-conservativity.
 Suppose now that $m$ is mono and that $\Phi(m)$ is iso. Then $m\in \Mc^\bot$ because the inequality $\exists_m (\top) \leq \top$ becomes an equality in ${\bf E_2}$, so that $\exists_m (\top) \simeq \top$.
 For the same reason, $m\times 1_Z$ is open for all $Z$ because $\Phi(m)\times 1_{\Phi(Z)}$ is open being an
 iso. Thus $m$ is a brittle isomorphism and so it is iso since $({\bf E_1}, \cM_1)$ is saturated.
\end{proof}

As an immediate corollary, we can strengthen our representation theorem:

\begin{theorem}\label{thm:main1} For every saturated small modal category $({\bf E}, \cM)$,
 the evaluation functor
 $$
 ev_{({\bf E}, \cM)}: ({\bf E}, \cM) \longrightarrow {\bf Rel^{G(E,\cM)}}
$$
 is a conservative embedding of modal categories.
\end{theorem}

\section{Quotients and Disjoint Unions}\label{sec:quotients}

Being a quasi-topos, $\cRel^{\bf G}$ is  a coherent category~\cite{elephant}.
This fact distinguishes it from  other semantic modal categories (e.g. topological spaces and preordered sets are not regular categories). 
In addition, in $\cRel^{\bf G}$ sums exist, are disjoint and pullback stable. However $\cRel^{\bf G}$ fails to be a pretopos because only equivalence relations which are regular monos are effective.
In this section we show how to enrich modal categories with quotients and disjoint unions, in order to eventually define the notion of a modal quasi-pretopos. To avoid unclear technical complications, we limit ourselves to \emph{Boolean} modal categories.

\begin{definition}\label{modal-cat-quotient}
	A \emph{Boolean modal category with quotients} is a Boolean modal category $(\cE,\Mc)$ equipped with a further distinguished class $\Qc$ of maps called \emph{quotients} and which satisfy the following conditions:
	\begin{enumerate}
		\item Each arrow $q \in \Qc$ is surjective in the internal logic, meaning that $\Qc \subseteq \Mc^\bot$.
		\item If $q : X\superto Y$ is a quotient, then for every $Z$ and for every $S\in \Sub_{\Mc}(Y\times Z)$ we have 
        \begin{equation}\label{eq:quotients1}
        \Diamond S \leq \exists_{q\times 1_Z} \Diamond (q\times 1_Z)^*S .
        \end{equation}
	\end{enumerate}
	A \emph{morphism} of Boolean modal categories with quotients is a modal functor which sends quotients to quotients.
\end{definition}

For instance, if $\Gb$ is a graph, then $\cRel^\Gb$ is equipped with a canonical quotient structure by declaring that $f : X\superto Y$ is a quotient when
\begin{itemize}
	\item $f_\alpha : X_\alpha\superto Y_\alpha$ is surjective for every $\alpha \in \Gb$ and
	\item if $k$ is an arrow of $\Gb$ and $s~k~t$ in $Y$, then there is $s', t' \in X$ with $s'~k~t'$ and $f(s') = s$ and $f(t') = t$.
\end{itemize}
Thus in $\cRel^\Gb$ (but \emph{not} in $\bf Top$) quotients turn out to be just regular epi.

Let us fix a small Boolean modal category with quotients $(\cE,\Mc,\Qc)$.
Recall from Section~\ref{sec:representation} the definition of a modal transformation between two coherent models $M$, $N$ of 
$(\cE,\Mc,\Qc)$: it is a family of relations 
 $R=\{R(X) \subseteq M(X) \times N(X)\}_{X\in\cE}$ satisfying conditions~\eqref{eq2}-\eqref{eq4}. We say that $R$ \emph{respects quotients} whenever
\begin{equation}\label{ax:modrel:quot}
	a~R~b \implies \exists a',b' : q(a')=a ~\&~ q(b')=b ~\&~ a'~R~b' 
\end{equation}
for any quotient $q$.

To continue, we need to introduced $\lambda$-saturated models (this is a standard notion in model theory~\cite{CK} that we are going to reformulate in our context).
Given $X,Y \in \cE$, we will denote by $\pi_Y : X\times Y \to Y$ the canonical projection.
Let $M : \cE \superto \cSet$ be a coherent model; take a lex subfunctor $\mathcal{S}\subseteq M$ and some object $X$ in $\cE$. An \emph{$X$-pretype over ${\mathcal S}$} is a collection $\Theta$ of subsets $\Theta_a \subseteq \Sub_\Mc(X\times Z)$ 
indexed by $a\in {\mathcal S}(Z)$
such that:
\begin{enumerate}
        \item[{\rm (PT0)}] $\Theta_a$ is a  \emph{prefilter}, i.e a non empty subset of $\Sub_\Mc(X\times Z)$ such that for each $A, B \in \Theta_a$, there is $C\in \Theta_a$ such that $C \leq A \land B$. Thus, the filter generated by $\Theta_a$ coincides with the up-set generated by $\Theta_a$, denoted ${\uparrow}\Theta_a$.
	\item[{\rm (PT1)}] If $(a',a)\in {\mathcal S}(Z'\times Z)$
         and $A\in \Theta_a$ then $(1_X\times\pi_Z)^* A\in {\uparrow}\Theta_{(a',a)}$.
	\item[\rm{(PT2)}] If $A\in \Sub_\Mc(Z)$, $\pi_Z:X\times Z \lra Z$ is the projection, and $a\in  \lbrack\!\lbrack A\rbrack\!\rbrack_M$, then $\pi_Z^*A \in {\uparrow} \Theta_a$.
\end{enumerate}

 We say that $c \in M(X)$ \emph{realizes} $\Theta$ if 
 for all $Z$, $a\in \mathcal{S}(Z)$ and $A\in \Theta_a$, we have that $(c,a)\in \lbrack\!\lbrack A\rbrack\!\rbrack_M $.
  We say that $\Theta$ is \emph{consistent} when ${\uparrow}\Theta_a$ never contains $\bot$. Given a regular cardinal $\lambda$, we say that 
  $M$ is \emph{$\lambda$-saturated} iff for every lex subfunctor $\mathcal{S}$ such that the cardinality of $\sum_{Z\in \cE} \mathcal{S}(Z)$ is less than $\lambda$,  every $M$-\emph{consistent} $X$-pretype over $\mathcal S$ is realized in $M$.

\begin{remark}
Given a pretype $\Theta$,
to facilitate the comparison with the definition from model theory textbooks,  it might be useful to write  $A\in \Theta_a$ as $A(X,a)\ \in \Theta$. Thus the elements of $\Theta$ can be viewed as formulae in the expanded language with constants from the support of $\mathcal S$. Seen in this way, condition (PT1) becomes tautological
and condition (PT2) just says that if $A(a)$ is true in $M$ then it belongs to $\Theta$. Similarly, $c$ realizes $\Theta$ iff for all 
$A(X,a)\ \in \Theta$ we have that $A(c,a)$ is true in $M$.
The notation $A(X,a)\ \in \Theta$ is more transparent,  notice however that it should be clear that
this notation does not refer to an  expansion of  the \emph{modal} language with constants. Rather we have pairs given by formulae in the original modal language and constants naming elements from the model. As a consequence, if $A$ is $\Diamond A'$, then in $\Diamond A'(X,a)$
the `constant' $a$ stays \emph{outside} the modal operator. In other terms, we expanded not the modal language but the classical first-order language used to encode the modal language, see Section~\ref{sec:theories} below for details.
\end{remark}

\begin{remark} 
If (PT2) is satisfied, in (PT0) there is no need to ask that $\Theta_a$ is not empty ($\pi_Z^*\top$ will be in $\Theta_a$ anyway).
  In a positive context, the axiom (PT2) has to be transformed into an equivalence: $a\in  \lbrack\!\lbrack A\rbrack\!\rbrack_M$ iff $\pi_Z^*A \in {\uparrow} \Theta_a$.
  In a Boolean context, the left-to-right direction is sufficient because then 
  $a\not\in  \lbrack\!\lbrack A\rbrack\!\rbrack_M$
   implies
   $\lnot\pi_Z^*A \in {\uparrow} \Theta_a$ and if also
   $\pi_Z^*A \in {\uparrow} \Theta_a$, then $\Theta$ is in fact inconsistent.
\end{remark}

\begin{remark}
    It can be shown that (PT1) can be equivalently stated using arrows which are not necessarily projections as follows: if $f:Z'\lra Z$ is such that $f_M(a')=a$
         and $A\in \Theta_a$ then $(1_X\times f)^* A\in {\uparrow}\Theta_{a'}$.
     This suggests a slightly different (but substantially equivalent) definition of a pretype, where one takes as $\Theta$ the filtered colimit of the 
     $\Theta_a$ (here the filtered colimit is taken over the opposite of the category of elements of $\mathcal{S}$, which is co-filtered because $\mathcal S$ is lex).
\end{remark}

In the following we will be interested in $\card{\,\cE}^+$-saturated models, where $\card{\,\cE}^+$ is the successor cardinal of the cardinality $\card{\,\cE}$ of the set of arrows of $\cE$.
Since applying  ultrapower modulo a suitable ultrafilter to a model  $M$ makes it
$\card{\,\cE}^+$-saturated~\cite[Thm 6.1.4 and 6.1.8]{CK}, it is clear that there are enough $\card{\,\cE}^+$-saturated models, e.g. in the sense that Proposition~\ref{prop:real} applies not only to plain coherent models but also to
$\card{\,\cE}^+$-saturated ones.

\begin{lemma}\label{gen4}
	Let $M,N$ be two $\card{\,\cE}^+$-saturated coherent models of the Boolean modal category with quotients $(\cE, \Mc,\Qc)$ and let $R\subseteq M\times N$ such that $\card{\,R} \leq \card{\,\cE}$. Suppose that $R$ satisfies \eqref{eq2} and \eqref{eq4}. Let $q : X\twoheadrightarrow Y$ be a quotient in $\cE$. Let $(a,b) \in Y_M \times Y_N$ such that $a\,R\,b$. Then $R$ is contained in some $R' \subseteq M\times N$ satisfying \eqref{eq2} and \eqref{eq4} and containing some $(a',b') \in X_M\times X_N$ such that $q(a') = a$ and $q(b') = b$. Moreover, $\card{\,R'} \leq \card{\,\cE}$.
\end{lemma}

\begin{proof}
We find first $a'$, $b'$ and finally $R'$.

{\bf Construction of $a'$.}
Let $\mathcal S$ be a lex subfunctor of $M$ such that $\sum_{Z\in \cE} \mathcal{S}(Z)$
has cardinality at most $\card{\,\cE}$ and  comprises all $c$ such that there is $c'$ with $cR c'$.  
Given  $f:Z\lra V$, 
$A\in \Sub_\Mc(Y\times Z)$ and  $B \in \Sub_\Mc(Y\times V)$ we write
	\[ F(f,A,B) = (q \times 1_Z)^*(A) \land (1_X\times f)^*\Diamond(q\times 1_V)^*(B) \text{.} \]
For $c\in \mathcal{S}(Z)$, let $\Theta_c$ be the set of the subobjects $S\in \Sub_\Mc(X\times Z)$ such that there exist 
$f:Z\lra V$, 
$d\in V_N$, $A\in \Sub_\Mc(Y\times Z)$, $B\in \Sub_\Mc(Y\times V)$  such that $f(c)Rd$, 
$S=F(f,A,B)$,
$(a,c)\in  \lbrack\!\lbrack A\rbrack\!\rbrack_M$ and  $(b,d)\in\lbrack\!\lbrack B\rbrack\!\rbrack_N$.
 This defines an $X$-pretype $\Theta$ over $\mathcal{S}$ because: 
 \begin{enumerate}
		\item $\Theta_c$ is a prefilter on $\Sub_\Mc(X\times Z)$ and (PT0) is satisfied: in fact, if we have $f_i:Z\lra V_i$, 
$d_i\in N_{V_i}$, $A_i\in \Sub_\Mc(Y\times Z)$, $B_i\in \Sub_\Mc(Y\times V_i)$  such that $f_i(c)Rd_i$, $S_i= F(f_i, A_i,B_i)$,
$(a,c)\in  \lbrack\!\lbrack A_i\rbrack\!\rbrack_M$ and  $(b,d_i)\in\lbrack\!\lbrack B_i\rbrack\!\rbrack_N$ (for $i=1,2)$, then taking 
$V:=V_1\times V_2$, $f:=(f_1,f_2)$, 
$$~A:=A_1\wedge A_2,~ B:=(\pi_{Y\times V_1})^*B_1\wedge (\pi_{Y\times V_2})^*B_2,$$ 
we have by continuity of pullbacks along projections
$$
 F(f,A,B) \leq 
 F(f_1,A_1,B_1) \wedge 
 F(f_2,A_2,B_2)
$$
and moreover
 $f(c)=(f_1(c),f_2(c))R(d_1,d_2)$ by \eqref{eq2}, 
$(a,c)\in  \lbrack\!\lbrack A\rbrack\!\rbrack_M$ and  $(b,d_1,d_2)\in\lbrack\!\lbrack B\rbrack\!\rbrack_N$.
\item If $(c',c)\in {\mathcal S}(Z'\times Z)$
         and $F(f,A,B)\in \Theta_c$ with $f:Z\lra V$ and the required data,
         we have that $(1_X \times \pi_Z)^* F(f,A,B) = F(\pi_Z f , \pi_{Y\times Z}^* A,B)\in \Theta_{c,c'}$ with the obvious data, so that (PT1) is satisfied.
\item If $C \in \Sub_\Mc(Z)$ and $c\in  \lbrack\!\lbrack C\rbrack\!\rbrack_M$, then $\pi_Z^*C \in {\uparrow} \Theta_c$: to see this, take  $f: Z\lra {\bf 1}$, $A:=\pi_Z^*C$ and $B=\top$.
           Thus (PT2) is satisfied.
	\end{enumerate}
The consistency of $\Theta$ is established as follows: if we  have (under the above mentioned  data)
$$
 F(f,A,B) =  (q \times 1_Z)^*(A) \land (1_X\times f)^*\Diamond(q\times 1_V)^*(B) \leq \bot
$$
then applying $\exists_{q\times 1}$ to this inequality, then using the Frobenius condition, the Beck--Chevalley condition and finally Definition~\ref{modal-cat-quotient}, we get
$$
A \land (1_Y\times f)^*\Diamond B
\leq \bot
$$
which is impossible since $(a,c)\in  \lbrack\!\lbrack A\rbrack\!\rbrack_M$,  $(b,d)\in\lbrack\!\lbrack B\rbrack\!\rbrack_N$, $aRb$, $f(c) R d$  and so $(a,f(c))\in  \lbrack\!\lbrack \Diamond B\rbrack\!\rbrack_M$, because $R$ satisfies 
\eqref{eq2} and \eqref{eq4}.

Thus $\Theta$ is realized in $M$ by some $a'\in M(X)$.
Taking in particular $f$ to be the identity arrow and $A$ to be the equality over $Y$ in the above data, we see that \emph{$ q(a')=a$ and that  
whenever we have $c R d$ and $(b,d)\in \lbrack\!\lbrack B\rbrack\!\rbrack_N$ we have also $(a',c)\in \lbrack\!\lbrack \Diamond (q\times 1)^* B\rbrack\!\rbrack_M$}.

{\bf Construction of $b'$.}
Let now $\mathcal S'$ be a lex subfunctor of $N$ such that $\sum_{Z\in \cE} \mathcal{S}(Z)$
has cardinality at most $\card{\,\cE}$ and  comprises all $d$ such that there is $c$ with $cR d$.   Given $f:Z\lra V$ and 
$A\in \Sub_\Mc(X\times V)$, $B\in \Sub_\Mc(Y\times Z)$ we write
	\[ G(f,A,B) = (1_X \times f)^*(A) \land (q\times 1_Z)^*(B) \text{.} \]
For $d\in \mathcal{S}'(Z)$, let $\Psi_d$ be the set of the subobjects $S\in \Sub_\Mc(X\times Z)$ such that there exist 
$f:Z\lra V$, 
$c\in V_M$, $A\in \Sub_\Mc(X\times V)$, $B\in \Sub_\Mc(Y\times Z)$  such that $cRf(d)$, 
$S=G(f,A,B)$,
$(a',c)\in  \lbrack\!\lbrack \Box A\rbrack\!\rbrack_M$ and  $(b,d)\in\lbrack\!\lbrack B\rbrack\!\rbrack_N$.
This defines an $X$-pretype $\Psi$ over $\mathcal{S}'$ because: 
 \begin{enumerate}
		\item    $\Psi_d$ is a prefilter on $\Sub_\Mc(X\times Z)$ and (PT0) is satisfied: in fact, if we have $f_i:Z\lra V_i$, 
$c_i\in M_{V_i}$, $A_i\in \Sub_\Mc(X\times V_i)$, $B_i\in \Sub_\Mc(Y\times Z)$  such that $c_iRf_i(d)$, 
$S_i=G(f_i,A_i,B_i)$,
$(a',c_i)\in  \lbrack\!\lbrack \Box A_i\rbrack\!\rbrack_M$ and  $(b,d)\in\lbrack\!\lbrack B_i\rbrack\!\rbrack_N$ (for $i=1,2$), then taking 
$f:=(f_1,f_2)$,
$$~A:=(\pi_{X\times V_1})^*A_1\wedge (\pi_{X\times V_2})^*A_2,~ B:=B_1\wedge B_2,$$ 
we have $G(f,A,B)\leq G(f_1,A_1,B_1) \wedge G(f_2,A_2,B_2)$
and moreover
 $(c_1,c_2)Rf(d)$, $(b,d)\in\lbrack\!\lbrack B\rbrack\!\rbrack_N$
 and finally $(a',c_1,c_2)\in  \lbrack\!\lbrack \Box A\rbrack\!\rbrack_M$
because 
$$
(\pi_{X\times V_1})^* \Box A_1\wedge (\pi_{X\times V_2})^*\Box A_2
\leq \Box ((\pi_{X\times V_1})^*A_1\wedge (\pi_{X\times V_2})^*A_2)
$$
by continuity.
 \item If $(d',d)\in {\mathcal S}(Z'\times Z)$
         and $G(f,A,B)\in \Psi_d$ with $f:Z\lra V$ and the required data, then
         $(1_X\times\pi_Z)^*G(f,A,B) = G(\pi_Z f,A,\pi_{Y\times Z}^* B)\in \Psi_{d,d'}$ with the obvious data, so that (PT1) is satisfied.
\item If $D \in \Sub_\Mc(Z)$ and $d\in  \lbrack\!\lbrack D\rbrack\!\rbrack_M$, then $\pi_Z^*D \in {\uparrow} \Psi_d$: to see this, take $f : Z \to \bf 1$, $c$ the unique element of $M_{\bf 1}$, $A=\top$ and $B=\pi_Z^*D$. Thus (PT2) is satisfied.
 \end{enumerate}       
The consistency of $\Psi$ is established as follows: if we  have (under the above mentioned  data)
\[  (1_X \times f)^*(A) \land (q\times 1_Z)^*(B) \leq \bot \]
then applying $\exists_{1_X\times f}$ and using the Frobenius  and the Beck-Chevalley conditions, we get
$$
A\wedge (q\times 1_V)^*\exists_{1_Y\times f} B\leq \bot
$$
and hence 
$$\Box A\wedge \Diamond (q\times 1_V)^*\exists_{1_Y\times f} B\leq \bot~~.$$
From $(b,d)\in\lbrack\!\lbrack B\rbrack\!\rbrack_N$ we get 
$(b,f(d))\in\lbrack\!\lbrack \exists_{1_Y\times f} B\rbrack\!\rbrack_N$ and from 
$cRf(d)$ and the construction of $a'$, we obtain 
$(a',c)\in  \lbrack\!\lbrack \Diamond (q\times 1_V)^*\exists_{1_Y\times f} B\rbrack\!\rbrack_M$. This, together with $(a',c)\in  \lbrack\!\lbrack \Box A\rbrack\!\rbrack_M$  yields a contradiction.

Thus $\Psi$ is consistent and realized in $N$ by some $b'$.  
Taking in particular $B$ to be the equality over $Y$ and $f$ to be the identity in the above data, we see that \emph{$ q(b')=b$ and
$(b',d) \in \lbrack\!\lbrack A\rbrack\!\rbrack_N$ implies $(a',c)\in \lbrack\!\lbrack \Diamond A\rbrack\!\rbrack_M$ whenever $cRd$}
(remember that $\Diamond=\neg \Box \neg$).

{\bf Construction of $R'$}.
Let $R'$ be the family of relations defined as follows, for every $Z$ and 
$c\in Z_M$, $d\in Z_N$:

\begin{center}
$c  R' d$ iff there exist $g:X\times Z'\lra Z, c'\in Z'_M, d'\in Z'_N$ such that $c' R d'$ and $c=g(a',c'), d=g(b',d')$.
\end{center}

\noindent
It is immediate to see that $R'$ satisfies conditions~\eqref{eq2} and \eqref{eq4}, so it is the desired $R'$.
\end{proof}

\begin{lemma}\label{gen5}
	Let $M,N$ be two $\card{\,\cE}^+$-saturated coherent models of a Boolean modal category with quotients $(\cE, \Mc,\Qc)$ and let $R_0\subseteq M\times N$ such that $\card{\,R_0} \leq \card{\,\cE}$. Suppose that $R_0$ satisfies \eqref{eq2} and \eqref{eq4}. Then $R_0$ is contained in some modal transformation  $R' \subseteq M\times N$ satisfying also \eqref{ax:modrel:quot}.
\end{lemma}

\begin{proof}
    Pick a well ordering of the triples $(a,b,q)$ (where 
    $q:X\lra Y$ is in $\Qc$, $a\in Y_M$, $b\in Y_N$ and $aRb$)
    by some ordinal $\kappa$ (notice that the cardinality of $\kappa$ does not exceed $\card{\,\cE}$). We define a sequence $(R_\xi \subseteq M\times N)_{\xi\leq\kappa}$ by:
	\begin{itemize}
		\item $R_0$ is given;
		\item $R_{\xi+1}$ is obtained by applying Lemma~\ref{gen4} to $R_\xi$ and the $\xi$-th element of our well ordering;
		\item $R_\lambda = \bigcup_{\xi<\lambda} R_\xi$ when $\lambda$ is a limit ordinal.
	\end{itemize}
	By induction, we have $\card{\,R_\xi} \leq \card{\,\cE}$ for all $\xi$. We then define $R'_1$ as $\tilde{R_\kappa}$, i.e. we apply  the construction of Lemma~\ref{lem:gen1} to ${R_\kappa}$. By referring to the construction presented in Lemma~\ref{lem:gen1}, we see that $\card{\,R'_1} \leq \card{\,\cE}$. Since $R_\kappa$ satisfies ~\eqref{eq2} and \eqref{eq4}, we obtain that $R'_1$ is a modal transformation satisfying 
    all conditions \eqref{eq2}-\eqref{eq4} and
    also \eqref{ax:modrel:quot}, but the latter only when $(a,b)$ is taken in $R_0 \subseteq R'_1$.
    To get the desired $R'$ it is now sufficient to repeat $\omega$-times the above construction leading from $R_0$ to $R'_1$, and take the union.
\end{proof}

\begin{theorem}\label{thm:repr_quotients}
	If $(\cE,\Mc,\Qc)$ is a small Boolean modal category with quotients, then there is a graph $\Gb$ and a morphism $\cE \superto \cRel^\Gb$ of Boolean modal categories with quotients which is $\Mc$-conservative.
\end{theorem}

\begin{proof}
    Let $\lambda=\card{\,\cE}^+$ be  the successor of the cardinal $\card{\,\cE}$.
	Let $\Gb_\lambda(\cE)$ be the graph of coherent $\lambda$-saturated models of the f-coherent category $(\cE,\Mc)$ and modal transformations which respect quotients. We will show that the canonical evaluation functor 
    $$ev : \cE \superto \cRel^{\Gb_\lambda(\cE)}$$ 
    is a morphism of Boolean modal categories. We know that it is $\Mc$-con\-ser\-va\-tive, since there are enough $\lambda$-saturated models
    (see the extension of Proposition~\ref{prop:real} to $\lambda$-saturated models).
    
    	The only non-immediate thing that we must check is that whenever $M \in \Gb_\lambda(\cE)$, for each $S \subseteq X$ in $\cE$ and each $a  \in \lbrack\!\lbrack \Diamond S\rbrack\!\rbrack_M$ there is some $N \in \Gb_\lambda(\cE)$, a modal transformation $R \subseteq M\times N$ which respects quotients and some $b \in \lbrack\!\lbrack  S\rbrack\!\rbrack_N$  such that $(a,b) \in R$. We start by building $N \in \Gb_\lambda(\cE)$ and $b \in X_N$ such that $b \in  \lbrack\!\lbrack  S\rbrack\!\rbrack_N$ and for any $S' \subseteq X$ such that $b \in  \lbrack\!\lbrack  S'\rbrack\!\rbrack_N$ we have $a \in \lbrack\!\lbrack \Diamond S'\rbrack\!\rbrack_M$. The reasoning is exactly the same as in the case without quotients. Lemma~\ref{gen5} now shows that the pair $(a,b) \in X_M \times X_N$ is contained in a modal transformation $R \subseteq M\times N$ respecting quotients.
\end{proof}

\begin{definition}\label{def:disjoint_unions}
In a modal category $(\cE,\Mc)$, a \emph{disjoint union} of two objects $A,B$ is an object $A+B$ endowed with two $\Mc$-subobjects
$$
\iota_A: A\hookrightarrow A+B, \qquad \iota_B: B\hookrightarrow A+B
$$
which are the Boolean complement of each other and such that 
\begin{equation}\label{eq:djunion}
\Diamond A\leq A~~~~and~~~~\Diamond B\leq B
\text{,}
\end{equation}
where the $\Diamond$ operator is taken inside of $A+B$.
\end{definition}

In  $\cRel^{\bf G}$ and in $\bf Top$ disjoint unions exist for every $A, B$ and are computed as in $\bf Set$.

\begin{proposition}
The disjoint union injections $\iota_A: A\hookrightarrow A+B, \iota_B: B\hookrightarrow A+B$ are open, meaning that we have $\iota_A^* \Diamond S = \Diamond \iota_A^* S$ for all $S\in \Sub_\Mc(A+B)$ and similarly for $B$.
\end{proposition}
\begin{proof}
    For $S\in \Sub_\Mc(A+B)$, we have $S=S_1\vee S_2$, with $S_1\leq A$ and $S_2\leq B$.
    Then $\iota_A^* \Diamond S=\iota_A^* \Diamond S_1 \vee \iota_A^* \Diamond S_2=\iota_A^* \Diamond S_1$
    by~\eqref{eq:djunion}.
    On the other hand $\Diamond \iota_A^* S= \iota_A^*\Diamond \exists_{\iota_A}\iota_A^* S=\iota_A^* \Diamond S_1$,
    by~\eqref{eq:subspace}. The argument showing openness of $\iota_B$ is the same.
\end{proof}

We now introduce Boolean modal quasi-pretoposes, but we warn the reader that the axioms for a Boolean modal quasi-pretopos acquire full transparent meaning only in presence of saturation conditions (as we show below).

\begin{definition}
 A \emph{Boolean modal quasi-pretopos} is a Boolean modal category with quotients $(\cE, \cM, \Qc)$ satisfying the following additional properties:
 \begin{enumerate}[(i)]
   \item every pair of objects $A, B \in \cE$ has a disjoint union $A+B$;
   \item every $R\in \Sub_{\Mc}(X\times X)$ which is an equivalence relation is the kernel of some 
   quotient $q : X\superto Q$.
 \end{enumerate}
\end{definition}

A morphism of Boolean modal quasi-pretoposes is simply a morphism of modal categories with quotients. Note that such a morphism automatically sends disjoint unions to disjoint unions, since the conditions they must satisfy involve only coherent and modal operators. 

$\cRel^{\bf G}$ is a Boolean modal quasi-pretopos (but $\bf Top$ is not because the quotient topology may fail to satisfy~\eqref{eq:quotients1}).
In $\cRel^{\bf G}$, disjoint unions are even unique modulo isomorphisms (in fact, as we will see below in Theorem~\ref{thm:saturated-modal-pretopos}, this is a consequence of saturation). 
The representation theorem extends to Boolean modal quasi-pretoposes:

\begin{theorem}\label{thm:repr_pretoposes}
	If $(\cE,\Mc)$ is a small Boolean modal quasi-pretopos, then there is a graph $\Gb$ and a morphism $\cE \superto \cRel^\Gb$ of Boolean modal quasi-pretoposes which is $\Mc$-conservative.
\end{theorem}

\begin{proof}
    Immediate by Theorem~\ref{thm:repr_quotients}.
\end{proof}

We now analyze some remarkable features of Boolean modal quasi-pre\-to\-po\-ses, that can all be deduced from the above representation theorem.

\begin{proposition}
In a Boolean modal quasi-pretopos $(\cE, \Mc)$, the class $\Mc$ is the class of regular monos.
\end{proposition}

\begin{proof}
    Let $S\buildrel{s}\over \hookrightarrow X$ be in $\Mc$. If we let $\iota_1, \iota_2$ be the two injections $X\rightrightarrows X+X$, we have that the reflexive closure of $(s\iota_1, s\iota_2):S \hookrightarrow (X+X)^2$ is an equivalence relation belonging to $\Mc$. Take its 
    quotient $q: X+X\twoheadrightarrow Q$ and then the equalizer $E  \buildrel{e}\over \hookrightarrow X$ of $\iota_1q$ and $\iota_2q$.
    By the universal property, there is $k$ such that $ke=s$. Now $k\in \Mc$ as the first component of an arrow in $\Mc$ (see Lemma~\ref{lem:firstcomponent}). However, it is not difficult to see that in every $\Mc$-conservative embedding $\eta$ of $(\cE, \Mc)$ into a category of the kind $\cRel^{\bf G}$, we have that $\eta(k)$ is iso (this is because coproducts and quotients of equivalence relations are computed set-wise in $\cRel^{\bf G}$). Thus $k$ is iso by Theorem~\ref{thm:repr_pretoposes} and so $s$ is an equalizer.
\end{proof}

In the hypothesis of saturation, more can be proved, because Theorem~\ref{thm:repr_pretoposes} supplies a conservative embedding (see Proposition~\ref{prop:conservativity}).

\begin{theorem}\label{thm:saturated-modal-pretopos}
    Let $(\cE, \Mc)$ be a saturated Boolean modal quasi-pretopos. Then:
    \begin{enumerate}[(i)]
   \item disjoint unions are coproducts;
   \item a zero object is an initial object;
   \item quotients coincide with regular epis;
   \item regular epis are pullback stable (so $\cE$ is a regular category).
   \end{enumerate}
\end{theorem}

\begin{proof}
  (Ad (i)).~
    Let 
    $$
    A\buildrel{\iota_A}\over \lra A+B \buildrel{\iota_B}\over\longleftarrow B
    $$
    be a disjoint union and let
    $$
    A\buildrel{f}\over \lra X \buildrel{g}\over\longleftarrow B
    $$
    be further arrows as displayed. We must find a unique 
    $h:A+B\lra X$ such that $h\circ\iota_A=f$ and $g\circ\iota_B=g$.
    Uniqueness comes from the fact that the equalizer
    $E\hookrightarrow A+B$ of two such $h$'s must be equal to $\top$, because it is easy to see, by first order reasoning, that 
    $E\wedge A=A$ and $E\wedge B=B$. For existence, consider the relation $R
    \buildrel{r}\over{\hookrightarrow} 
    (A+B)\times X$ given by
    $$
    (\exists_{\iota_A\times 1_X}(f\times 1_X)^* =_X)
    \vee
    (\exists_{\iota_B\times 1_X}(g\times 1_X)^* =_X)~.
    $$
    In the internal language of doctrines, this can be expressed by the formula 
    $$
    \exists a \,(\iota_A(a)=z \wedge f(a)=x) \vee 
    \exists b \,(\iota_B(b)=z \wedge f(b)=x)
    $$
    (with free variables $z$ of type $A+B$ and $x$ of type $X$).
    By first-order reasoning it is possible to see that
    the composite arrow
    $$
    k:R\buildrel{r}\over{\hookrightarrow} (A+B)\times X \buildrel{\pi_{A+B}}\over\lra A+B
    $$
    is both in $\Mc^\bot$ (i.e. that $\exists_k(\top)=\top$) and mono
    (i.e. that $k(z_1)=k(z_2)\to z_1=z_2$ holds) using the internal language. Moreover, applying the characterization of disjoint unions in relational $\bf G$-sets, in every conservative embedding $\eta$ of $(\cE, \Mc)$ into a category of the kind $\cRel^{\bf G}$, $\eta(k)$ turns out to be an isomorphism. By conservativity, $k$ is an isomorphism and then 
    $$
    A+B\buildrel{k^{-1}}\over \lra R \buildrel{r}\over {\hookrightarrow} (A+B)\times X\buildrel{\pi_X}\over \lra X
    $$
    gives the required $h$.
    
    (Ad (ii)).~ Given a zero object $O$ and any object $X$, we must find a unique $h:O\lra X$. Uniqueness comes from the fact that the equalizer of any two such arrows must be the identity, because identity is the unique $\cM$-subobject of $O$. For existence, consider the projection $\pi_O:O\times X \lra O$: this is mono and becomes iso in any representation into relational $\bf G$-sets, so it is iso by conservativity. The inverse of $\pi_O$ composed with the projection into $X$ gives the required $h$. 

    (Ad (iii)).~Suppose we are given a quotient $q:X\twoheadrightarrow Y$. We show that $q$ is a regular epimorphism. Take $f:X\lra Z$ with $ker(q)\leq ker(f)$. We need to find a unique $h:Y\lra Z$ such that $h\circ q=f$. Uniqueness comes from the fact that $q$ is epi, as any arrow in $\Mc^\bot$
    (recall Proposition~\ref{prop:epis}). Consider the relation $R\buildrel{r}\over{\hookrightarrow} Y\times Z$ given by
    $$
    \exists_{\pi_{Y\times Z}}(\pi_{X\times Z}^* (q\times 1_Y)^* {=_Y} \land \pi_{X\times Z}^* (f\times 1_Z)^* {=_Z})
    $$
    In the internal language of doctrines, this can be expressed by the formula 
    $$
    \exists x\,  (q(x)= y\wedge f(x)=z)~~.
    $$
    By first-order reasoning it is possible to see that
    the composite arrow
    $$
    k:R\buildrel{r}\over{\hookrightarrow} Y\times Z \buildrel{\pi_Y}\over\lra Y
    $$
    is both in $\Mc^\bot$ 
    and mono.
    Moreover, applying the characterization of quotients in relational $\bf G$-sets, in every conservative embedding $\eta$ of $(\cE, \Mc)$ into a category of the kind $\cRel^{\bf G}$, $\eta(k)$ turns out to be an isomorphism. By conservativity, $k$ is an isomorphism and then 
    $$
    Y\buildrel{k^{-1}}\over \lra R \buildrel{r}\over {\hookrightarrow} Y\times Z\buildrel{\pi_X}\over \lra Z
    $$
    gives the required $h$.

    We have shown that any quotient $q$ is the coequalizer of its kerner $ker(q)$. Since any equivalence relation is of the form $ker(q)$, it follows that, reciprocally, any regular epimorphism is a quotient.
    
    (Ad (iv)).~We show that the pullback of a regular epi $q: X\lra Q$ along any arrow is a regular epi. Let us factorize such a pullback
    $q'$  as
  $$
  X'\buildrel{q''}\over \lra Q' \buildrel{m}\over \lra Q''
  $$
  where $q''$ is a regular epi  ($q''$ is the quotient, aka the  coequalizer, of the kernel of $q'$) and $m$ is mono by the lemma below. In a conservative representation $\eta$ of $(\cE, \cM)$ into a relational $\bf G$-sets category, it turns out $\eta(m)$ is an iso because  relational $\bf G$-sets are  a regular category, hence $m$ itself is iso by conservativity.
\end{proof}

\begin{lemma}
 In an  f-regular category $(\cE,\Mc)$, suppose that $f: X\lra Y$ factors as 
 $X\buildrel{q}\over\lra Q\buildrel{m}\over \lra Y$, where $q$ is such that $q\in \Mc^\bot$
 and $ker(f)\leq ker(q)$ (this is certainly the case if $q$ is the coequalizer of $ker(f)$, recall Proposition~\ref{prop:epis}). Then $m$ is mono.
\end{lemma}
\begin{proof}
 This can be established using first order reasoning via the internal language.  
 Alternatively, make a diagram chasing, using the fact that the class $\Mc^\bot$ is pullback-stable and is contained in the class of epis by  Proposition~\ref{prop:epis}. 
 %
 %
 %
 %
 %
\end{proof}

 From the above facts, it follows that  in a saturated Boolean modal quasi-pretopos
  every arrow $f:X\lra Y$ factors as 
$$
X\buildrel{q}\over{\lra} X' \buildrel{\iota}\over{\lra} Y'\buildrel{m}\over{\lra} Y
$$
where $q$ is a regular epi, $m$ is a regular mono, and $\iota$ is both mono and epi; 
moreover such factorization is pullback stable.


\section{Modal first-order theories}\label{sec:theories}

We present a calculus for first order modal logic and we show how to associate a modal category with each theory so as to prove a completeness theorem for modal first order theories.
For simplicity, we supply a full classical logic calculus in Hilbert style with primitive logical operators $\rightarrow, \bot, \forall, \Box$: readers interested in relevant fragments (like the coherent fragment) \emph{can easily formulate by themselves the corresponding calculi} using e.g. suitable sequents.

The main problem we have to face in building a syntactic calculus corresponding to our modal categories lies in the fact that modal operators are not pullback stable, they are only `half-stable' (see the continuity condition in Definition~\ref{def:main}). This problem has been somewhat known for a long time in the first-order modal logic literature (especially in the literature oriented towards philosophical applications), although of course it was not formulated in this way. One of the proposed solutions in such literature was the adoption of a restricted form of $\lambda$-abstraction~\cite{ST68}: this is the solution we want to imitate below, in order to obtain a calculus which is more flexible than the calculus presented in~\cite{GhiModalitaCategorie1990,handbook}.
The idea is to combine abstraction operators with a variant of the `de-modalization' technique employed in some literature~\cite{awodey_aiml,Ginvariance}. The outcome of this is a full  encoding of our first-order modal theories into classical first-order theories.

We need to use formulae-in-context, as commonly adopted in the categorical logic literature to handle possibly empty domains (in particular, we shall follow rather closely the formalism of~\cite[D1]{elephant} when building our language and our syntactic categories).\footnote{ 
The only little difference with~\cite[D1]{elephant} is that we build recursively the context of a formula together with the formula itself. The consequence of our choice is that variables cannot have  in a formula both a free occurrence and an occurrence bound by a quantifier (this is a desirable feature, giving that formulae are taken modulo $\alpha$-conversion).
On the other hand, there is a major difference with respect to the language introduced in~\cite{GhiModalitaCategorie1990,handbook}, because our contexts are finite list of variables and not lists of sorts: the present choice makes the language much less rigid.
}

We fix a many sorted first-order signature (with equality) $\Sigma$: sorts are indicated with letters $X,Y, \dots$, atomic predicates with letters $P,Q, \dots$ and function symbols with letters $f, g\,\dots$ To each predicate or function symbol an \emph{arity domain} (i.e. a list of sorts) is associated; function symbol are also assigned a \emph{codomain} sort. We have a countably infinite supply of variables for every sort $x_1^X, x_2^X, \dots, y_1^Y, y_2^Y, \dots$ (the sorts of the variables may be omitted if confusion does not arise). A finite \emph{list of variables without repetitions}
is called a \emph{context}: contexts are indicated with letters $\ux, \uy, \dots$ (the list of the sorts of the variables in a context is often left implicit in the notation). 

\emph{Terms-in-contexts} are defined as follows:
\begin{enumerate}[(i)]
 \item
  $x^X: \ux$ is a term in context of type $X$ if $x^X\in \ux$;
  \item
  if $t_1:\ux, \dots, t_n:\ux$ are terms in contexts of types
$X_1,\dots, X_n$ respectively, then $f(t_1, \dots, t_n):\ux$ is a term in context of type $X$ if the function symbol $f$ has $X$ as codomain sort and $X_1\cdots X_n$ as arity domain.
\end{enumerate}

We now define formulae-in-context and predicates by mutual recursion.

\emph{Formulae-in-contexts} are defined as follows:
\begin{enumerate}[(i)]
 \item
 if $t_1:\ux, \dots, t_n:\ux$ are terms in contexts of types
$X_1,\dots, X_n$ respectively, then $P(t_1, \dots, t_n):\ux$ is a formula in context,  if the predicate  $P$ has  $X_1\cdots X_n$ as arity domain;
\item $\bot:\ux$ is a formula in context,  for every context $\ux$;
\item if $\varphi_1:\ux$ and $\varphi_2:\ux$ are formulae in context, so is $(\varphi_1\to \varphi_2):\ux$;
\item if $x^X\in \ux$ and $\varphi:\ux$ is a formula  in context, so is
$(\forall x^X \varphi): \ux\setminus\{x^X\}$;
\end{enumerate}

\emph{Predicates} are so defined:
\begin{enumerate}
\item[{\rm (a)}] atomic predicates $P$ with domain arity $X_1\cdots X_n$
are predicates with domain arity $X_1\cdots X_n$;
\item[{\rm (v)}] if  $\varphi:\ux$ (where $\ux= x_1^{X_1},\dots, x_n^{X_n}$) is a formula  in context, then
$$
\Box\, \{
 \ux \mid\varphi\}
$$
is a predicate with domain arity $X_1\cdots X_n$.
\end{enumerate}
We underline that the abstraction operator applies to the context as a whole and not to single variables.
Below,  if $\ux= x_1^{X_1},\dots, x_n^{X_n}$ and if $t_1:\uy, \dots, t_n:\uy$ are terms in contexts of types
$X_1,\dots, X_n$ respectively,
we may write simply
$$(\Box \varphi)(t_1, \dots, t_n) : \uy$$
instead of $
\Box\, \{
 \ux \mid\varphi\} (t_1, \dots, t_n)
$. We may also write   $\Box \varphi :\ux$ for 
$(\Box \varphi)(\ux):\ux$.

We treat our formulae modulo $\alpha$-conversion, meaning that alphabetic variants will be considered the same formula (notice that both $\forall$ and $\{ \ux \mid -\}$ are variable binders).
Substitution can be inductively defined as expected: 
if $y_1^{Y_1}\cdots y_m^{Y_m}$ are the variables of the context $\uy$
of a formula in context $\varphi:\uy$
and if $u_1:\ux, \dots, u_m:\ux$ are terms in context of types
$Y_1,\dots, Y_m$ respectively, we define the formula in context
$$
\varphi[u_1/y_1^{Y_1}, \dots, u_m/y_m^{Y_m}]: \ux
$$
as follows (we assume that the bounded variables occurring in $\varphi$ are disjoint from $\ux$ --- if it is not so, an alphabetic variant of $\varphi$ is taken before applying the definition below):
\begin{enumerate}[(i)]
 \item for every predicate $P$ (atomic or not), we define  $$P(t_1, \dots, t_n)[u_1/y_1^{Y_1}, \dots, u_m/y_m^{Y_m}]:\ux$$
 as the formula in context $P(t'_1, \dots, t'_n):\ux$, where $t'_i$ (for $1\leq i\leq n$) is obtained from $t_i$ by replacing in it each occurrence of $y_1^{Y_1}, \dots, y_m^{Y_m}$
 by $u_1, \dots, u_m$, respectively;
  \item $\bot [u_1/y_1^{Y_1}, \dots, u_m/y_m^{Y_m}]:\ux$ is the formula in context $\bot:\ux$;
   \item $(\varphi_1\to \varphi_2)[u_1/y_1^{Y_1}, \dots, u_m/y_m^{Y_m}]:\ux$ is
   the formula in context given by  $(\varphi_1[u_1/y_1^{Y_1}, \dots, u_m/y_m^{Y_m}]\to \varphi_1[u_1/y_1^{Y_1}, \dots, u_m/y_m^{Y_m}]):\ux$;
  \item  $(\forall x^X \varphi) [u_1/y_1^{Y_1}, \dots, u_m/y_m^{Y_m}]: \ux$
  is the formula in context given by $\forall x^X \varphi [u_1/y_1^{Y_1}, \dots, u_m/y_m^{Y_m}, x^X/x^X]: \ux$.
\end{enumerate}

\noindent
Below we shall usually write $
\varphi[u_1/y_1^{Y_1}, \dots, u_m/y_m^{Y_m}]: \ux
$  as
$$
\varphi[u_1, \dots, u_m]: \ux
$$
(or just as $\varphi[\uu]$ if $\uu=u_1, \dots, u_m$)
for simplicity, if confusion does not arise.
As a special case, 
for a predicate $P$ (atomic or not),
we may use the notations $P(t_1, \dots, t_n):\ux$ and 
$P[t_1, \dots, t_n]:\ux$ interchangeably.
Finally, $
\varphi[u_1/y_1^{Y_1}, \dots, u_m/y_m^{Y_m}]: \ux
$ can be abbreviated as $\varphi[u_i/y_i^{Y_i}]$ if $u_j$ is equal to $y_j$ for all $j\neq i$.



%
\begin{table}[htbp]
\hrule
~~
\\
\caption{
 } \label{tab:deduction} \hrule

\begin{center}
\begin{tabular}{l}
\begin{tabular}{lr}
{\bf Axiom Schemata}
\\
$ \varphi:\ux$ & ($Taut$) \\
(provided $\varphi$ is an instance of a tautology)
\\
\\
$\Box (\varphi\to \psi) \to (\Box \varphi \to \Box \varphi):\ux$ &($\Box Dis$)
\\
\\
$\forall y^Y \varphi\to \varphi[t/y^Y]: \ux$ & ($\forall$-$Ex$)
\\
\\
$t=t: \ux$ & ($Refl$)
\\
\\
$t_1=t_2 \to  (\varphi[t_1/y^Y]\to \varphi[t_2/y^Y]):\ux$ &
($Repl$)
\\
\\
$(\Box \varphi)[t_1,\dots, t_n]\to \Box (\varphi[t_1,\dots, t_n]):\ux$ &($Cont$)
\\
\\
{\bf Inference Rules}
\\
\hbox{
\begin{tabular}{lll}
 $\psi:\ux$ & &$\psi\to \varphi:\ux$
 \\
\hline  &$\varphi:\ux$&
\\
\end{tabular}}
& ($MP$)
\\
\\
\hbox{
\begin{tabular}{l}
 $\varphi:\ux$
 \\
\hline  $\Box\varphi:\ux$
\\
\end{tabular}}
& ($Nec$)
\\
\\
\hbox{
\begin{tabular}{l}
 $\varphi[\ux']\to \psi~~:\ux$ 
 \\
\hline  $\varphi\to\forall y^Y\psi:\ux'$\\
(where $\ux'=\ux\setminus\{y^Y\}$)
\\
\end{tabular}}
& ($\forall$-$In$)
\\
\\
\hbox{
\begin{tabular}{l}
 $\varphi: \ux$
\\
\hline  $\varphi[\ut]:\uy$
\\
\end{tabular}}
& ($Inst$)
\\
 ~(where $\ut:\uy$ is a tuple of terms of  types $\ux$).
\\
\end{tabular}

\end{tabular}

\end{center}

\vskip 2mm
 \hrule
\end{table}

Axioms and rules for the calculus are specified in Table~\ref{tab:deduction}.
Few observations are in order: $(Taut)$,$(\Box Dis)$,$(MP)$,$(Nec)$ are the standard axiomatization of the modal system $K$, as reported in modal logic textbook like~\cite{CZ}. On the other hand, $(\forall$-$Ex)$, $(\forall$-$In),(Refl),(Rep)$ are the well-known Hilbert-style axiomatization of classical first-order logic with equality.
The only specific axiom of our framework in the continuity axiom $(Cont)$. 
Rule $(Inst)$ is admissible in the pure calculus and is reported here in analogy to~\cite[D1]{elephant}.

Given a \emph{modal theory } $T$ (i.e.\ a set of formulae in context) and a formula in context $\varphi:\ux$, we write
$$
\vdash^{\ux}_T \varphi
$$
(and we say that $\varphi:\ux$ is \emph{derivable} from $T$)
to mean that there is a derivation of $\varphi:\ux$ that uses the above axioms and rules together with the formulae of $T$ as extra axioms. Notations like $\varphi_1, \dots, \varphi_n \vdash^{\ux}_T \varphi$ mean $\vdash^{\ux}_T \varphi_1\wedge \cdots\wedge \varphi_n \rightarrow \varphi$.

An \emph{evaluation} of the signature $\Sigma$ into a modal category $(\bf  E, \cM)$ is a map $\mathcal{I}$ that associates
\begin{itemize}
 \item with every sort $X$ of $\Sigma$ an object $\mathcal{I}(X)$ of $\bf E$;
 \item with every atomic predicate $P$ of $\Sigma$ of domain $X_1\cdots X_n$ an $\cM$-subobject $\mathcal{I}(P)$
of $\mathcal{I}(X_1)\times \cdots \times \mathcal{I}(X_n)$;
\item with every function symbol $f$ of $\Sigma$ of arity domain $X_1\cdots X_n$ and codomain $X$ an arrow $\mathcal{I}(f): \mathcal{I}(X_1)\times \cdots \mathcal{I}(X_n)\longrightarrow \mathcal{I}(X)$ in $\bf E$.
\end{itemize}
By recursion, it is clear how to extend $\mathcal I$ to terms and formulae in context, so as to associate with a formula in context $\varphi:\ux$ (where $\ux=x_1^{X_1},\dots, x_n^{X_n}$) an $\cM$-subobject
$\mathcal{I}(\varphi)\hookrightarrow
\mathcal{I}(X_1)\times \cdots \mathcal{I}(X_n)$. The evaluation is called a \emph{model} of a modal theory $T$ in $(\bf E,\cM)$ iff for every $\varphi:\ux$ in $T$ we have that
$\top=\mathcal{I}(\varphi)$.

The following categorical completeness theorem links our calculus with modal categories:

\begin{theorem}
We have $\vdash_T^{\ux} \varphi$ iff  $\top=\mathcal{I}(\varphi)$ holds for every model $\mathcal I$ of $T$ in a Boolean modal category $(\bf E,\cM)$.
\end{theorem}

\begin{proof}

One side of the theorem below is just an induction on the length of derivations, whereas the other side requires the construction of the modal category $({\bf E}_T, \cM_T)$ canonically associated with a modal theory $T$. We shall merely sketch here such a construction, 
relying on the formalism taken  from~\cite{elephant}. Although the detailed computations involved in the construction are indeed rather annoying,  the construction by itself is interesting because it shows how to build the suitable factorization system, taking inspiration from the conditions of Definition~\ref{def:main}.


  We take as objects the equivalence classes of formulae in context
  $$
  \{ \varphi: \ux\}
  $$
  where we consider $\varphi:\ux$ and $\varphi':\ux'$ to be equivalent if $\varphi':\ux'$ is a \emph{renaming} of $\varphi:\ux$ (this means that there is an alphabetic variant $\tilde \varphi:\ux$ of $\varphi:\ux$ and a variable bijection $\ux\longmapsto \ux'$ such that $\varphi':\ux'$ is equal to
  $\tilde \varphi  [\ux'/\ux] : \ux'$). In defining morphisms from $\{ \varphi: \ux\}$ to $\{ \psi: \uy\}$ we may assume that the contexts $\ux$ and $\uy$ are disjoint. A formula in context $R:\ux,\uy$ such that
  \begin{eqnarray*}
  & R\vdash^{\ux,\uy}_T \varphi\wedge \psi \\
  & R[\ux, \uy], R[\ux, \uy']\vdash^{\ux,\uy,\uy'}_T \uy=\uy' \\
  & \varphi\vdash^{\ux}_T \exists \uy R
  \end{eqnarray*}
  is called a \emph{functional relation} between $\varphi: \ux$ and $\psi: \uy$ (here $\uy=\uy'$ means the conjunction of the componentwise equalities). We take as arrows between $\{ \varphi: \ux\}$ and $\{ \psi: \uy\}$ the equivalence classes (wrt to provable equivalence in $T$) of such functional relations that satisfy in addition the following requirement for all formulae in context $\theta: \uy ,\uw$ with $\uw\cap \ux=\emptyset$:
  \begin{equation}\label{eq:pcont}
   \varphi[\ux]\wedge \Diamonda \exists \uy (R[\ux,\uy]\wedge \theta[\uy,\uw])\vdash^{\ux,\uw}_T
   \exists \uy (R[\ux,\uy]\wedge (\Diamonda \theta)[\uy,\uw])~~.
  \end{equation}
  This formula is the transcription in our language of the continuity condition~\eqref{eq:Rcont0} for a relation $R$ in a modal category 
   (the extra conjunction with $\varphi[\ux]$ appearing in~\eqref{eq:pcont} is due to the fact that $\varphi$ itself is meant to be a subobject of the object represented by the context $\ux$).

  Following~\cite{elephant}, the composite of two morphisms
  \[\begin{tikzcd}
      \{\varphi:\ux\} \ar[r, "{\{R\}}"] & \{\psi:\uy\} \ar[r, "{\{S\}}"] & \{\chi:\uz\}
  \end{tikzcd}\]
  is $\{\exists \uy (R[\ux,\uy]\wedge S[\uy,\uz])\}$ (notice that the context of the formulae representing an arrow can be desumed from the domain and the codomain), whereas the identity of $\{\varphi:\ux\}$ is
  \[
  \begin{tikzcd}
    \{\varphi:\ux\} \ar[r, "{\{\varphi[\ux] \wedge \ux = \ux'\}}"] &[4em] \{\varphi[\ux'/\ux]:\ux'\}\text{.}
  \end{tikzcd}
  \]
  Associativity and identity conditions can be checked as in~\cite{elephant},
  but the continuity condition~\eqref{eq:pcont} has to be checked in addition (we omit the sufficiently straightforward syntactic details).

  Removing the continuity condition~\eqref{eq:pcont} and repeating the considerations in~\cite{elephant}, we obtain a coherent Boolean category (that we call ${\bf C}_T$) and an obvious faithful inclusion functor 
  ${\bf E}_T \lra {\bf C}_T$: the first fact to check is  that this functor 
  preserves finite limits.
  
  In ${\bf C}_T$, the terminal object is  $\{\top:\; \}$ (its context is empty) and $\{ \varphi: \ux\}\buildrel{\{\varphi\}}\over\lra \{\top:\; \}$ is the unique map from $\{ \varphi: \ux\}$ to the terminal object.
  The product of $\{ \varphi: \ux\}$ and $\{ \psi: \uy\}$  is $\{ \varphi[\ux]\wedge \psi[\uy]: \ux,\uy\}$ and the projection maps
  are
  \[\begin{tikzcd}[column sep=6.5em]
      \{\varphi[\ux'/\ux]:\ux'\} &
      \{ \varphi[\ux]\wedge \psi[\uy]: \ux,\uy\}
      \ar[l,"{\{\varphi[\ux]\wedge \psi[\uy]\wedge \ux=\ux'\}}"']
      \ar[r,"{\{\varphi[\ux]\wedge \psi[\uy]\wedge \uy=\uy'\}}"]
      &
      \{\psi[\uy'/\uy]:\uy'\}
  \end{tikzcd}\]
  The equalizer of $\{\varphi:\ux\}\buildrel{\{R\}}\over\lra\{\psi:\uy\}$ and of $\{\varphi:\ux\}\buildrel{\{S\}}\over\lra\{\psi:\uy\}$
  is
  \[\begin{tikzcd}[column sep=10em]
      \{\exists \uy (R[\ux'/\ux, \uy]\wedge S[\ux'/\ux, \uy]): \ux'\}
      \ar[r,"{\{\exists \uy (R[\ux, \uy]\wedge S[\ux, \uy])\wedge \ux'=\ux\}}"]
      &
      \{\varphi:\ux\}
  \end{tikzcd}\]
  Since all involved maps (including the ones needed for the universal properties of limits) satisfy the continuity condition~\eqref{eq:pcont},
  we have that indeed ${\bf E}_T$ is a lex subcategory of ${\bf C}_T$. 

  We now have to identify an adequate stable factorization system in
  ~$ {\bf E}_T$. In the end, we shall see that the above faithful inclusion preserves the chosen factorization systems (which is the 
  strong epi/mono factorization system in ${\bf C}_T$), but some subtleties arise, due to the fact that iso's are not reflected
  (in fact, an iso in ${\bf C}_T$ is iso in ${\bf E}_T$ too iff \emph{both} 
  it and its inverse satisfy the continuity condition~\eqref{eq:pcont}).
  
  We take as the right class $\mathcal{E}_T$ in ${\bf C}_T$ the set of the maps $\{ \varphi: \ux\}\buildrel{\{R\}}\over \lra\{ \psi: \uy\}$
  satisfying the condition $\psi\vdash^{\uy} \exists \ux\, \varphi$.
  These are the maps that are strong epi in ${\bf C}_T$. Now, like in  ${\bf C}_T$, every map factors in ${\bf E}_T$ as a map in $\mathcal{E}_T$ followed by a `canonical subobject', i.e. by a map of the kind
  \[\begin{tikzcd}
    \{\varphi[\ux'/\ux]: \ux'\} \ar[r, "{\varphi[\ux]\wedge \ux'=\ux}"] &[4em] \{\psi: \uy\}
  \end{tikzcd}\]
  where $\varphi\vdash_T^{\ux} \psi$. Here comes however the main difference: in ${\bf C}_T$, every monic arrow is isomorphic to a canonical subobject (see~\cite{elephant} for a proof), whereas this is not true anymore in ${\bf E}_T$ because we lack sufficiently many isomorphisms. This means that the closure under isomorphisms of the class of canonical subobjects in ${\bf E}_T$ is a more restricted class (still orthogonal to $\mathcal{E}_T$, though), which we are going to identify.

  Recall~\cite{elephant} that an arrow $\{ \varphi: \ux\}\buildrel{\{M\}}\over \lra\{ \psi: \uy\}$ is monic (in ${\bf C}_T$ and also in ${\bf E}_T$) iff its satisfies the condition
  $$
  M(\ux_1,\uy),  M(\ux_2,\uy) \vdash^{\ux_1, \ux_2,\uy} \ux_1 = \ux_2~~. 
  $$
  We let $\cM_T$ be the class of monic arrows of ${\bf E}_T$
  satisfying the further condition (below we use notations like 
  $ \zeta_1=^{\uz}_T \zeta_2$ 
  to mean `$\zeta_1\vdash_T^{\uz} \zeta_2$ and 
  $\zeta_2\vdash_T^{\uz} \zeta_1$')  
  \begin{equation}\label{eq:monoET}
   \varphi[\ux] \wedge \Diamonda \theta~ =^{\ux,\uw}_T
   ~\exists \uy\,(M[\ux,\uy]\wedge (\Diamonda \exists \ux' ( M[\ux',\uy] \wedge \theta[\ux',\uw]))[\uy,\uw])  \end{equation}
   for every formula  $\theta:\ux,\uw$ such that 
   $\theta\vdash^{\ux,\uw}_T \varphi[\ux]$.
Notice that~\eqref{eq:monoET} is nothing but  a             `translation' of condition~\eqref{eq:subspace}. 

The following facts can now be checked via suitable syntactic proofs:
\begin{enumerate}[(i)]
 \item $\cM_T$ is orthogonal to $\mathcal{E}_T$;
 \item $\cM_T$ is closed under composition and contains iso's;
 \item if $\{M\}, \{M'\}$ are  in $\cM_T$ and 
 $\{ M\}=\{ M'\}\circ \{R\}$, then $\{R\}$ is in $\cM_T$ too.
\end{enumerate}
\[\begin{tikzcd}[column sep=0.5em,every label/.append style = {font = \normalsize}]
\{\varphi:\ux\} \ar[rd, "{\{M\}}"'] \ar[rr, "{\{R\}}"] & & \{\chi:\uz\} \ar[ld, "{\{M'\}}"] \\[1em]
& \{\psi:\uy\} &
\end{tikzcd}\]
As a consequence we have that every arrow $\{M\}$ in $\cM_T$ is isomorphic to a canonical subobject: this is  because the first component of the factorization of $\{M\}$ (as an arrow in $\cE$ followed by a canonical subobject) is in $\mathcal{E}_T\cap \cM_T$, so it is an iso.

To complete the construction of $({\bf E}_T, \cM_T)$, we need to introduce the modal operators. 
This is done as follows: if $\{ M\}\simeq\{\psi[\ux'/\ux]:\ux'\}$ is an $\cM_T$-subobject of $\{\varphi: \ux\}$ 
we let 
$$
\Diamond \{M\}~:=~\{(\varphi \wedge \Diamonda \psi)[\ux'/\ux]:\ux'\}~~.
$$
The conditions of Definition~\ref{def:main} are easily seen to be satisfied by construction. 

It is now straightforward to define a canonical evaluation $\mathcal{I}_T$ such that we have $\mathcal{I}_T(\varphi:\ux)=\{\varphi:\ux\}$ for every formula $\varphi:\ux$. The claim of the theorem follows immediately.
\end{proof}

\section{Conclusion and Further Developments}\label{sec:conclusions}

We proposed an approach to quantified modal logic via logical categories: our modal categories are lex categories endowed with an adequate factorization system $(\cE, \cM)$, where the semilattices of $\cM$-subobject are distributive lattices with a modal operator satisfying the axioms of modal logic $K$. These modal distributive lattices interact with the factorization system via continuity conditions for inverse images and subspace conditions for $\cM$-subobjects. The interaction becomes tighter when (mutually equivalent) saturation axioms for $\cM$-subobjects, definable functional relations and definable isomophisms are assumed. To justify our framework we proved  representation theorems in semantic modal categories which are quasi-toposes and realize  a counterpart semantics.
Still, we believe that our understanding of modal categories needs more investigations: in this section, we mention some problems and some directions for future research.
\begin{enumerate}[1.]
\item The relationship between modal categories and modal first order theories needs to be studied further. The reader might have noticed that the modal category $({\bf E}_T,\cM_T)$ built in Section~\ref{sec:theories} is a saturated modal category. Should this fact give strong motivation towards the adoption of the saturation principles in the very definition of a modal category? Not quite. The saturation principles are essentially higher order principles: they analyze conditions under which a definable subobject/functional relation/bijection is promoted to become a real subspace/continuous function/isomorphism. The conditions are \emph{relative to the available $\cM$-subobjects} (we would in fact say: ``to the available current language''), so they might fail in a larger environment where more $\cM$-subobjects arise. Thus, there is no reason why the internal models of a modal first order theory $T$ into a modal category $({\bf E}, \cM)$ should correspond to 
modal functors 
from $({\bf E}_T,\cM_T)$ to $({\bf E}, \cM)$. Such property is likely to be enjoyed by a smaller modal category than $({\bf E}_T,\cM_T)$, namely the `minimum modal subcategory' of $({\bf E}_T,\cM_T)$ comprising the interpretation of terms and of predicates of the language of $T$ as arrows and as $\cM$-subobjects, respectively. Such a smaller category will have less isomorphisms, so the closure under isomorphisms of the `canonical subobjects' class will be different than in $({\bf E}_T,\cM_T)$.  
Thus, modal categories might carry more information than modal first order theories, in the sense that they also declare explicitly that certain subobjects/functional relations/bijections have to be considered as real subspaces/continuous functions/isomorphisms. Such extra information is partially undetermined in a modal first order theory
and it can be maximized or minimized going though appropriate modal subcategories of $({\bf E}_T,\cM_T)$. 


%

\item Adding $S4$ axioms~\eqref{eq:closure} is not as harmless as one may expect. From the semantic side, $\cRel^{\bf G}$ should be replaced by a relational presheaf category $\bf Rel^C$~\cite{louvain,handbook,Niefield}: here $\bf C$ is a category (not just a graph), objects are lax set-valued functors and arrows are the same as in $\bf Rel^C$. This category is not as well-behaved as $\cRel^{\bf G}$ (to see it, notice that $\bf Rel^C$ is $\bf POr$, when $\bf C$ is the singleton category, so it is not a quasi-topos). A completeness theorem for modal first order-theories has been proved in~\cite{louvain,handbook}, but the extension to modal categories does not look easy. The reason is that the
preservation of the $\cM$ class is responsible for the introduction of the reflection condition~\eqref{eq:refl} in the notion of a modal transformation and this reflection condition is not directly compatible with the lax composability that modal transformations should 
satisfy in the new $S4$ environment. Once again, these extra difficulties show that moving from modal first order theories to the more expressive framework of modal categories is far from trivial.

\item Trying to obtain sufficient conditions for a representation theorem into a (power of) $\bf Top$ is even more challenging. The conditions~\eqref{eq:closure} and ~\eqref{eq:products} are sufficient to prove a completeness theorem for a first order modal theory in a purely relational signature~\cite{GhiModalitaCategorie1990}. Dropping the limitation to a relational signature might be at hand, but moving from modal first order theories to modal categories faces the same difficulties analyzed above for the $S4$ axioms and relational presheaves. 

\item The mechanism for introducing modalities of the present paper does not require full first order logic, it operates already in the fragment of coherent logic. However, it is possible to weaken this fragment further. In fact we need existential quantifiers only in order to formulate the subspace condition~\eqref{eq:subspace}, but a closer inspection to this condition reveals that it 
requires only direct images along monos,
so condition~\eqref{eq:subspace} makes sense in the essentially algebraic fragment of lex categories. Disjunctions are needed to formulate $K$-axioms, but $K$-axioms might be weakened themselves. This is important in the description logics framework, where lighter logics are introduced. In particular, the description logic $\mathcal{EL}$ requires from the algebraic side just $\wedge$-semilattices with a monotone operator $\Diamond$.  $\mathcal{EL}$ plays an important role in the applications because it is computationally tractable and relevant industrial size ontologies can be formalized in $\mathcal{EL}$. It is evident that our framework offers a natural candidate for the formalization of a predicate extension of $\mathcal{EL}$ in the purely essentially algebraic context.

\item As an opposite research direction, one should explore extensions to higher order modal logic: the fact that our main semantic frameworks (namely relational $\bf G$-sets) are quasi-toposes makes this perspective viable.
\end{enumerate}

\bibliographystyle{plain}

\bibliography{biblio,biblio2}

\def\cprime{$'$}
\begin{thebibliography}{10}

\bibitem{awodey_rsl}
Steve Awodey and Kohei Kishida.
\newblock Topology and modality: the topological interpretation of first-order
  modal logic.
\newblock {\em Rev. Symb. Log.}, 1(2):146--166, 2008.

\bibitem{awodey_aiml}
Steve Awodey and Kohei Kishida.
\newblock Topological completeness of first-order modal logic.
\newblock In {\em Advances in modal logic. {V}ol. 9}, pages 1--17. Coll. Publ.,
  London, 2012.

\bibitem{description_logics}
Franz Baader, Ian Horrocks, Carsten Lutz, and Ulrike Sattler.
\newblock {\em An Introduction to Description Logic}.
\newblock Cambridge University Press, 2017.

\bibitem{rosalie}
Guram Bezhanishvili, Nick Bezhanishvili, and Rosalie Iemhoff.
\newblock Stable canonical rules.
\newblock {\em J. Symb. Log.}, 81(1):284--315, 2016.

\bibitem{BorHandbookCategoricalAlgebra1994}
Francis Borceux.
\newblock {\em Handbook of Categorical Algebra: Volume 1: Basic Category
  Theory}, volume~1 of {\em Encyclopedia of Mathematics and its Applications}.
\newblock Cambridge University Press, Cambridge, 1994.

\bibitem{mu_calculus}
Julian~C. Bradfield and Colin Stirling.
\newblock Modal mu-calculi.
\newblock In Patrick Blackburn, J.~F. A.~K. van Benthem, and Frank Wolter,
  editors, {\em Handbook of Modal Logic}, volume~3 of {\em Studies in logic and
  practical reasoning}, pages 721--756. North-Holland, 2007.

\bibitem{handbook}
Torben Bra\"uner and Silvio Ghilardi.
\newblock First-order modal logic.
\newblock In {\em Handbook of modal logic}, volume~3 of {\em Stud. Log. Pract.
  Reason.}, pages 549--620. Elsevier B. V., Amsterdam, 2007.

\bibitem{CZ}
Alexander Chagrov and Michael Zakharyaschev.
\newblock {\em Modal logic}, volume~35 of {\em Oxford Logic Guides}.
\newblock The Clarendon Press, Oxford University Press, New York, 1997.
\newblock Oxford Science Publications.

\bibitem{CK}
Chen-Chung Chang and Jerome~H. Keisler.
\newblock {\em Model Theory}.
\newblock North-Holland, Amsterdam-London, third edition, 1990.

\bibitem{pdl}
M.~Fischer and R.~Ladner.
\newblock Propositional dynamic logic of regular programs.
\newblock {\em Journal of Computer and System Sciences}, 18:194–211, 1979.

\bibitem{many_dimensional}
D.~M. Gabbay, A.~Kurucz, F.~Wolter, and M.~Zakharyaschev.
\newblock {\em Many-dimensional modal logics: theory and applications}, volume
  148 of {\em Studies in Logic and the Foundations of Mathematics}.
\newblock North-Holland Publishing Co., Amsterdam, 2003.

\bibitem{Val}
D.~M. Gabbay, V.~B. Shehtman, and D.~P. Skvortsov.
\newblock {\em Quantification in nonclassical logic. {V}ol. 1}, volume 153 of
  {\em Studies in Logic and the Foundations of Mathematics}.
\newblock Elsevier B. V., Amsterdam, 2009.

\bibitem{Gadducci}
Fabio Gadducci, Andrea Laretto, and Davide Trotta.
\newblock Specification and verification of a linear-time temporal logic for
  graph transformation.
\newblock In Maribel Fern{\'{a}}ndez and Christopher~M. Poskitt, editors, {\em
  Graph Transformation - 16th International Conference, {ICGT} 2023,
  Proceedings}, volume 13961 of {\em Lecture Notes in Computer Science}, pages
  22--42. Springer, 2023.

\bibitem{viareggio}
S.~Ghilardi and G.~Meloni.
\newblock Relational and topological semantics for temporal and modal
  predicative logic.
\newblock In {\em Nuovi problemi della logica e della scienza}, volume~II, page
  59–77. CLUEB Bologna, 1991.

\bibitem{louvain}
S.~Ghilardi and G.~C. Meloni.
\newblock Modal and tense predicate logic: models in presheaves and categorical
  conceptualization.
\newblock In {\em Categorical algebra and its applications
  ({L}ouvain-{L}a-{N}euve, 1987)}, volume 1348 of {\em Lecture Notes in Math.},
  pages 130--142. Springer, Berlin, 1988.

\bibitem{GhiModalitaCategorie1990}
Silvio Ghilardi.
\newblock {\em Modalit{\`a} e categorie}.
\newblock PhD thesis, Universit{\`a} degli Studi di Milano, 1990.

\bibitem{incompleteness}
Silvio Ghilardi.
\newblock Incompleteness results in {K}ripke semantics.
\newblock {\em J. Symbolic Logic}, 56(2):517--538, 1991.

\bibitem{Ginvariance}
Silvio Ghilardi.
\newblock The invariance modality.
\newblock In {\em V.A. Jankov on Non-Classivcal Logics, History and Philosophy
  of Mathematics}, volume~24 of {\em Outstanding Contributions to Logic}, pages
  165--175. Springer, 2022.

\bibitem{GMJSL}
Silvio Ghilardi and Giancarlo Meloni.
\newblock Relational and partial variable sets and basic predicate logic.
\newblock {\em J. Symbolic Logic}, 61(3):843--872, 1996.

\bibitem{GhilardiZawadowski2011}
Silvio Ghilardi and M.~Zawadowski.
\newblock {\em Sheaves, Games, and Model Completions: A Categorical Approach to
  Nonclassical Propositional Logics}.
\newblock Springer Publishing Company, Incorporated, 1st edition, 2011.

\bibitem{shirasu}
Shirasu H.
\newblock Duality in superintuitionistic and modal predicate logics.
\newblock In {\em Advances in Modal Logic I}, number~87 in CSLI Lecture Notes,
  pages 223--236. CSLI Publ., Stanford, CA, 1998.

\bibitem{hazen}
A.~Hazen.
\newblock Counterpart-theoretic semantics for modal logic.
\newblock {\em J. Philos.}, 76(6):319–--338, 1979.

\bibitem{temporal_logic}
Ian~M. Hodkinson and Mark Reynolds.
\newblock Temporal logic.
\newblock In Patrick Blackburn, J.~F. A.~K. van Benthem, and Frank Wolter,
  editors, {\em Handbook of Modal Logic}, volume~3 of {\em Studies in logic and
  practical reasoning}, pages 655--720. North-Holland, 2007.

\bibitem{elephant}
Peter~T. Johnstone.
\newblock {\em Sketches of an elephant: a topos theory compendium. {V}ol. 2},
  volume~44 of {\em Oxford Logic Guides}.
\newblock The Clarendon Press, Oxford University Press, Oxford, 2002.

\bibitem{hyperdoctrines}
William Lawvere.
\newblock Equality in hyperdoctrines and comprehension schema as an adjoint
  functor.
\newblock {\em Proceedings of the AMS Symposium on Pure Mathematics XVII},
  pages 1--14, 1970.

\bibitem{lewis}
D.K. Lewis.
\newblock Counterpart theory and quantified modal logic.
\newblock {\em J. Philos.}, 65(5):113–--126, 1968.

\bibitem{MaiRosQuotientCompletionFoundation2013}
Maria~Emilia Maietti and Giuseppe Rosolini.
\newblock Quotient completion for the foundation of constructive mathematics.
\newblock {\em Logica Universalis}, 7(3):371--402, September 2013.

\bibitem{makkai_reyes_apal}
M.~Makkai and G.~E. Reyes.
\newblock Completeness results for intuitionistic and modal logic in a
  categorical setting.
\newblock {\em Ann. Pure Appl. Logic}, 72(1):25--101, 1995.

\bibitem{MR}
Michael Makkai and Gonzalo~E. Reyes.
\newblock {\em First order categorical logic}, volume Vol. 611 of {\em Lecture
  Notes in Mathematics}.
\newblock Springer-Verlag, Berlin-New York, 1977.
\newblock Model-theoretical methods in the theory of topoi and related
  categories.

\bibitem{manna}
Zohar Manna and Amir Pnueli.
\newblock {\em Temporal Verification of Reactive Systems - Safety}.
\newblock Springer, 1995.

\bibitem{tarski}
J.~C.~C. McKinsey and A.~Tarski.
\newblock The algebra of topology.
\newblock {\em Annals of Mathematics}, 45:141–--191, 1944.

\bibitem{Niefield}
Susan Niefield.
\newblock Change of base for relational variable sets.
\newblock {\em Theory Appl. Categ.}, 12:No. 7, 248--261, 2004.

\bibitem{Niefield1}
Susan Niefield.
\newblock Lax presheaves and exponentiability.
\newblock {\em Theory Appl. Categ.}, 24:288--301, 2010.

\bibitem{pasquali}
Fabio Pasquali.
\newblock Remarks on the tripos to topos construction: comprehension,
  extensionality, quotients and functional-completeness.
\newblock {\em Appl. Categ. Structures}, 24(2):105--119, 2016.

\bibitem{reyes}
Gonzalo~E. Reyes.
\newblock A topos-theoretic approach to reference and modality.
\newblock {\em Notre Dame J. Formal Logic}, 32(3):359--391, 1991.

\bibitem{ST68}
Robert~C. Stalnaker and Richmond~H. Thomason.
\newblock Abstraction in first-order modal logic.
\newblock {\em Theoria}, 34(3):203--207, 1968.

\bibitem{vanMarDualityModelTheory2024}
Sam {van Gool} and J{\'e}r{\'e}mie Marqu{\`e}s.
\newblock On duality and model theory for polyadic spaces.
\newblock {\em Annals of Pure and Applied Logic}, 175(2):103388, February 2024.

\end{thebibliography}

\end{document}